\documentclass[showpacs,preprintnumbers,amsmath,amssymb]{revtex4}
\usepackage{graphicx}
\usepackage{dcolumn}
\usepackage{color}
\usepackage{bm}
\usepackage{float}
\usepackage{epstopdf}
\usepackage{hyperref}
\usepackage{cleveref}

\def\bea {\begin{eqnarray}}
\def\eea {\end{eqnarray}}

\def\be {\begin{equation}}
\def\ee {\end{equation}}

\begin{document}
\title{Calculation of Power Spectrum in the Little Bangs}
\author{Golam Sarwar$^{a,b}$, Sushant K. Singh$^{a,b}$,  Jan-e Alam$^{a,b}$. 
}
\medskip
\affiliation{$^a$ Variable Energy Cyclotron Centre, 1/AF, Bidhan Nagar,
Kolkata - 700064, INDIA, $^b$Homi Bhabha National Institute, Mumbai, India.
}
\begin{abstract}
The power spectrum of fluctuations in the momentum distributions of particles have been
estimated with optical Glauber and Monte-Carlo Glauber initial conditions  for relativistic
heavy ion collisions.  The evaluation procedure adopted in this work is analogous to 
the one used in the calculation of power spectrum in Cosmic Microwave Background Radiation (CMBR).
The power spectrum due to perturbations in the phase space distribution of the particles
has also been evaluated. The perturbation in phase space has
been evolved through the Boltzmann transport equation in an expanding quark gluon plasma (QGP) 
background. The expansion of the QGP has been treated within the purview of (3+1) dimensional relativistic hydrodynamics.  
We observe that the non-equlibrium  effects introduced as perturbations in the phase space 
distributions can be traced from the enhancement of the power spectrum as well as through its
variation with temperature which is distinctly different from the case of vanishing perturbation. 
A relation has been derived between the power spectrum and the flow harmonics.

\end{abstract}

\maketitle

\section{Introduction}
The renormalization group approach in Quantum Chromodynamics (QCD) predicts that 
at high density and temperature ($T$) hadronic matter undergoes a phase transition to
quark matter~\cite{collins} due to asymptotic freedom ~\cite{asymp1,asymp2} and Debye 
screeinging of colour charges. Calculations based on lattice QCD has predicted 
that the transition from hadronic to quark matter occurs in the temperature 
domain, $145\leq T$(MeV)$\leq 163$~\cite{bazavov,borsayni,karsch} at net negligible baryon density.
At Relativistic Heavy Ion Collider (RHIC)
and Large Hadron Collider (LHC) heavy nuclei are made to collide to  create 
quark matter or quark gluon plasma (QGP) - a state of matter that prevailed
in the micro-second old universe according to the cosmological Big Bang (BB) model. 
In this regard the production of QGP in nuclear collisions at relativistic
energies is dubbed as Little Bangs (LB).  
One of the compulsion to study the QCD phase transition in 
Heavy Ion Collisions at Relativistic Energies (HICRE) is to understand the
non-abelian gauge theory in medium and to understand the dynamics of 
similar transition in the early universe. This is especially 
important because the universe has undergone several other transitions {\it e.g.} 
Electroweak, GUT, etc,
but among these the QCD transition is the only one which is accessible through the presently 
available accelerator energy.  The study of the temperature fluctuation
in the cosmic microwave background radiation (CMBR) originated from the recombination era
(about 300,000 years after the BB)
has provided crucial knowledge in supports of BB model~\cite{HD,cmbr} and matter content
of the universe. The  polarization of photons 
due to Thomson scattering from the anisotropic 
decoupling surface (where the photon had suffered the last interactions) 
results in the non-zero quadrupole moment of the phase space distribution of the incident photon. 
This anisotropic fluctuations in density, for example, may be caused by the propagation of 
gravitational wave in the early universe.  
The  temperature fluctuation in the CMBR  is  introduced as a perturbation 
in the phase space distribution of photons. The evolution of this perturbation is studied
by using Boltzmann transport equation (BTE)~\cite{cmbr,cpma} in gravitational field.
The linear polarization due to the scattering is connected with the quadrupole
moment of the phase space distribution of photon. 

In this work we would like to perform
a theoretical analysis of LB following  procedure similar to the one used
in the analysis of CMBR. 
Study of fluctuation can be useful to characterize the state of the matter and also to put 
constraints on models~\cite{EvntFlc,IFC, DT,RL,SG,RLO,CnsMI1,CnsMI2,GJO,impCF}. 
Power spectrum in HICRE has been discussed
by several authors in Refs.~\cite{apm,sorensen,morphology,estrada}. In Ref.~\cite{apm} 
the root mean square of various flow harmonics has been calculated 
and shown strong similarities with the power spectrum of CMBR.  M\'ocsy and Sorensen~\cite{sorensen}
has extracted the power spectrum of the system produced in HICRE by using data on transverse momentum ($p_T$)
correlations.  In Ref.~\cite{estrada} data from ALICE collaboration has been used to estimate $p_T$ fluctuation
and subsequently expanded in Laplace series to estimate the power spectrum analogous to temperature
fluctuation in CMBR.  
In Ref.~\cite{morphology} relativistic heavy ion collision events have been generated by using HIJING~\cite{xnwang} 
and redistributed the produced particles to emulate flow effects to reproduce elliptic flow with required value.

As mentioned before the state of the matter, the QGP  created in HICRE imitates the condition that prevailed 
in the micro-second old universe. The space time evolution of the matter is 
governed by fluid dynamics for both BB and LB.  However, there are glaring differences too. 
For example, the relevant interactions, characteristic length and 
time scales in LB  and BB are very different, primarily because of the  pertinence of 
gravity in the BB.
In HICRE there are several sources of fluctuations.
Fluctuations in thermal variables have been suggested as  signals
of the critical end point in the QCD phase transition ~\cite{flc1,flc2,flc3}. Different magnitude of 
fluctuations in partonic and hadronic phases in the  net electric charge and baryon number
may shed light on the QCD phase transition in HICRE~\cite{asakawa}. Fluctuations in the ratio of
positively to negatively charged pions may be used as an indicator of QCD transition~\cite{jeon1} 
as well as for understanding the  chemical equilibrium in the system formed in HICRE~\cite{jeon}. 

In the present work we will study the power spectrum due to fluctuations 
in the initial energy density that may arise naturally due to the quantum fluctuation 
of the finite "lump-like" nucleons  within the colliding nuclei~\cite{flc4,qin}. These fluctuations evolve
hydrodynamically~\cite{pstaig1,pstaig2,RJ,kapusta,ripples,UW1,UW2,hanna}.
The bulk matter {\it i.e.} QGP created in HICRE with very high temperature and pressure will
expand relativistically against vacuum. This expansion is treated in the present work by
solving relativistic hydrodynamic equations in (3+1) dimensions with initial conditions 
derived from Optical as well as Monte-Carlo Glauber model~\cite{MCG}. The equation of state (EoS)
is taken from  QCD.  The system will revert to hadrons due to the cooling caused 
by the expansion.  In the hadronic phase the system may continue to expand hydrodynamically until
the mean free path of the constituents become too large to maintain equilibrium. When
the hadrons cease to interact, their momenta  get frozen and  hit the detector with that  
momenta. However, it has been shown that the chemical freeze-out of the hadrons 
takes place near the quark-hadron transition boundary, meaning that the system may
be out of chemical equilibrium in the hadronic phase and the evolution
of the hadronic phase can not be studied using hydrodynamics, it may require 
hybrid model approach (hydro+URQMD~\cite{rqmd}) which is beyond the scope 
of the present work. Therefore, we will  evaluate the power spectrum  of
QGP phase only in this work.  

We will also study the power spectrum of anisotropic
fluctuations in momentum space inflicted through the phase space distribution 
that may drive the system slightly away from equilibrium. The correlation that
survive the evolution can be  observed in the final state and which may be
connected to the initial state correlation~\cite{SG1,SG2,SG3,SG4,RLO,CnsMI1,CnsMI2,GJO}. 
The variation of the power spectrum with time will indicate
the dissipation of fluctuations created during the evolution.   The nature
of the variation may help in differentiating the fluctuations produced in
the initial state from those created afterward.   
The evolution of such anisotropic fluctuations 
is dictated by the Boltzaman transport equation (BTE).    
The propagation of the jets through QGP  may cause such fluctuations~\cite{DJ}. 
Therefore,  we intend to study the evolution of the fluctuations through 
BTE in a hydrodynamically expanding QGP background. 
The BTE is solved in relaxation time ($\tau_R$) approximation. 
$\tau_R$ has been taken from
calculations done by using hard thermal loop approximation in QCD.
In principle, $\tau_R$ is a function of temperature ($T$) and baryonic chemical 
potential ($\mu_B)$, however, as discussed below in the present case we need to consider
the $T$ dependence only. The change in $T$  due to expansion of the 
bulk is controlled by relativistic hydrodynamics. This change in $T$ affects evolution
of the fluctuation (solution of the BTE) through the relaxation time, indicating a direct coupling 
between the anisotropic fluctuation and bulk expanding background. 
The BTE has been solved with initial conditions containing spatial anisotropies to be specified later.

The initial energy density distribution, $\epsilon(\tau,x,y,\eta$) of the bulk matter created in HICRE can be estimated 
by using Glauber model. In the present work, both the Optical Glauber (OG) as well as the Monte-Carlo Glauber (MCG) 
models have been used to demonstrate the sensitivity of the results on initial conditions of the bulk matter.  The finite size
of the colliding nucleons with quantum fluctuations in the nuclear beams will create "lumpiness" in $\epsilon(\tau,x,y,\eta)$. 
This can be seen very clearly in $\epsilon(\tau,x,y,\eta)$ calculated using MCG. We study the
evolution of the these fluctuations using hydrodynamics
at the surfaces of constant $T$.
Power spectrum due to fluctuations caused by  phase space perturbations has also been estimated. 
In the present work we make an attempt to evaluate the fluctuations in HICRE 
in keeping close resemblance with analysis of temperature fluctuation in cosmic
microwave radiation (CMBR).  The power spectrum will be evaluated at  various stages of the evolving 
system to understand how it changes with time.

The paper is organized as follows. In the next section we will briefly discuss
the evolution of the quark gluon plasma within the framework of
relativistic hydrodynamics followed by discussions on the initial conditions and equation of state
used in this work in the successive subsections. 
The evolution of the fluctuations within the scope of 
BTE has been discussed in section III. The power spectrum has been evaluated in section IV.
Section V is devoted to present the results and section
VI is dedicated to summary and discussions. 

\section{Hydrodynamic evolution of the quark gluon plasma} 
The expansion of the  QGP in space and time can be described by applying relativistic 
hydrodynamics. The conservation of energy and momentum of the fluid is governed by the
equation: 
\be
\partial_{ \mu} T^{\mu \nu}=0
\label{hydroeq1}
\ee
where $T^{\mu\nu}=(\epsilon+P)u^\mu u^\nu - g^{\mu\nu}P$. Here $\epsilon$ is the
energy density, $P$ is the pressure, $u^\mu=\gamma(1,\vec{v})$ is the four 
velocity of the fluid and $\gamma=1/\sqrt{1-v^2}$. 
The conservation of
the net baryon number throughout the  evolution history is controlled by the 
equation:
\be
\partial_{ \mu}( n_B u^{\mu})=0
\label{hydroeq2}
\ee
where $n_B$ is the net baryon (baryon - antibaryon) density. 
However, in the present work we are interested in the system produced
in nuclear collisions at highest RHIC energies where
$n_B$ is negligibly small ($n_B$ will be even smaller at LHC collision
conditions) and hence $\mu_B\sim 0$. Therefore, we do not need to consider Eq.~\ref{hydroeq2}.
In the present work Eq.~\ref{hydroeq1} has been sloved numerically using standard
technique~\cite{hlle} in full (3+1) space-time dimension without assuming boost invariance along longitudinal
direction~\cite{bjorken} and cylindrical symmetry of the system.  The initial conditions 
and equation of state (EoS) used here are discussed briefly below.

\subsection{Initial conditions}
The initial conditions required to solve Eq.~\ref{hydroeq1} in (3+1) dimension are as follows:
The Cartesian components of initial flow velocities are: $v_x(\tau_0,x,y,z)=v_y(\tau_0,x,y,z)=0$ and
the initial energy density profile is taken as~\cite{hlle}:
\begin{equation}
 \varepsilon (\tau_0,x,y,\eta _s)= \varepsilon _{GM}(x,y) \, \theta (Y_b-|\eta _s |)\, 
\text{exp}\left[ -\theta (|\eta _s| -\Delta \eta)\frac{(|\eta _s| -\Delta \eta)^2}{\sigma _\eta ^2}\right]
 \end{equation}
 where $\varepsilon _{GM}(x,y)$ is obtained from OG or MCG model, having the following expression
 \begin{equation}
 \label{glauber_ed}
 \varepsilon _{GM}(x,y) = \varepsilon _0\left[ \frac{1-f}{2}n_{part}(x,y)+f n_{coll}(x,y)\right]
 \end{equation}
We have taken the value of the inelastic nucleon-nucleon cross section at RHIC energy as, $\sigma_{NN}=42$ mb
in evaluating the number of participants, $n_{part}$ and number of collisions, $n_{coll}$. 
In  MCG  model approach the energy density is deposited at discrete points, 
but for hydrodynamic evolution we need a continuous distribution of energy density. 
Therefore,  we use Gaussian smearing to get the energy density as:
\begin{equation}
\varepsilon _{GM}(x,y)=\frac{1}{2\pi \sigma ^2}\sum _{i} \, \varepsilon _{GM}(x_i,y_i)\, e^{-\frac{(x-x_i)^2+(y-y_i)^2}{2\sigma ^2}}
\end{equation}
 where $\varepsilon _{GM}(x_i,y_i)$ is obtained from Eq(\ref{glauber_ed}). To sample the
nucleons from nuclei (Au in this case), we use the following Woods-Saxon distribution
 $$\rho (r)=\frac{\rho _0}{1+e^{\frac{r-R}{\delta}}}$$
 
The values of  different  parameters appeared in the above expressions are tabulated below. 
 \begin{center}
 \begin{tabular}{|c|c|}
 \hline
  Parameter & Value \\
  \hline
  $\tau _0$ & 0.6 fm/c \\
  $Y_b$ & 5.3\\
  $\Delta \eta$ & 1.3 \\
  $\sigma _\eta $ & 2.1\\
  $\varepsilon _0$ & 7.7 Gev/$\text{fm}^3$\\
  $f$ & 0.14 \\
  $\sigma ^2$ & 0.16 \\
  $R$ & 6.37 \\
  $\delta$ & 0.535\\
  $\sigma _{NN}$ & 42 mb \\
 \hline
 \end{tabular}
 \end{center}
Table 1: Values of different parameters used in solving the hydrodynamical equations (see text for details).

\subsection{Equation of State (EoS)}
The EoS for the QGP and the hadrons have been constructed following the procedure outlined in Ref~\cite{albright}. 
We use excluded volume model ~\cite{ref_excvol} 
for hot hadrons and pQCD results ~\cite{vuorinen1,vuorinen2,mustafa,albright} for the QGP phase. For a smooth crossover, 
a switching function is used as in ~\cite{albright} and the parameters are adjusted so as to match the Lattice QCD results. 
A brief description of the model used is as follows. 
We choose volume of hadrons to be 
proportional to mass, $v_i=m_i/m_0$ as in ~\cite{albright}, where $m_0$ is a constant. 
We take $m_0=0.9$ for this work. The pressure of the hadronic medium is taken to be 
 \begin{align}
p_{HG}(T,\mu _B)&=\sum _{i=1} p_i^{id}(T,\tilde{\mu}_i) \label{HG_pressure}\\
\tilde{\mu}_i&= \mu _i-v_ip_{HG}
\end{align}
where $\mu _i= B\mu _B$ and $B$ is baryon number. $p_i^{id}$ denotes the ideal pressure of a relativistic 
gas comprised of $i^{th}$ resonance and $p_{HG}$ is the pressure after excluded volume correction is 
taken into account which is found by solving the above set of equations in a self-consistent way.  
 
The pressure of the QGP phase is taken as
 \begin{equation}
P=\frac{8\pi ^2}{45}T^4\left[ f_0+\left(\frac{\alpha_s}{\pi}\right)f_2
+\left(\frac{\alpha_s}{\pi}\right)^{3/2}f_3+\left(\frac{\alpha_s}{\pi}\right)^{2}f_4
+\left(\frac{\alpha_s}{\pi}\right)^{5/2}f_5+\left(\frac{\alpha_s}{\pi}\right)^3f_6\right]
\label{pressure}
\end{equation}
where the coefficients $f_n$'s are given in the appendix A. The coupling, $\alpha_s$ has been 
taken from ~\cite{pdg} calculated in three loop approximations. The pressure in the crossover 
region is taken to be
 \begin{equation}
  P(T,\mu)=S(T,\mu)P_{qgp}(T,\mu)+(1-S(T,\mu))P_h(T,\mu)
 \end{equation}
 where the switching function $S(T,\mu)$ is taken as
 \begin{align}
  S(T,\mu)&= \text{exp}\{ -\theta(T,\mu)\}\\
  \theta(T,\mu) &= \left[ \left( \frac{T}{T_0}\right)^r+ \left( \frac{\mu}{\mu_0}\right)^r\right]^{-1}
 \end{align}
 We take $T_0=165$ MeV, $\mu _0=3\pi T_0$ and $r=4$. With these parameter values we find 
a good agreement of our results with the lattice data~\cite{katz_lattice}. 
\section{Evolution of anisotropies and fluctuations}
In this section, we discuss the connection of $\delta f$, a small deviation in phase space 
distribution from its equilibrium value with the fluctuations in various 
thermodynamic quantities.  
The phase space distribution function, $f(\vec{x},\vec{p},t)$
of a system slightly away from equilibrium, at time $t$, position $\vec{x}$,
momentum $\vec{p}$ can be written as~\cite{eml}:
$f(x,p)=f_0(x,p)+\delta f(x,p)$ with $\delta f= f_0 \psi$. The evolution
of $\delta f$ is governed by the  BTE, $p^{\mu}\partial_{\mu}f=(p\cdot u)C[f]$~\cite{eml,degroot,cercignani}. 
For an expanding system under the relaxation time approximation  BTE reduces to the 
following (see ~\cite{sarwar} for details)~\cite{FRS1,FRS2,Hatta,expan2,expan3,expan4}):
\begin{equation}
\left( \frac{\partial}{\partial t}+\frac{\vec{p}}{p^0}\cdot\frac{\partial}{\partial \vec{x} }
+\frac{(p^0u_0-\vec{p}\cdot\vec{u})}{p^0 \tau_R(x)}\right)\delta f(x,p)
=-\left( \frac{\partial}{\partial t}+
\frac{\vec{p}}{p^0}\cdot \frac{\partial}{\partial \vec{x} }\right)f_0(x,p)
\label{bterelax}
\end{equation}
Eq.~\ref{bterelax} can be solved by using standard techniques used for partial differential equations~\cite{hlevine}. 
The solution is given by:
\begin{equation}
\delta f(x,p)=D(t,t_0)\left[ \delta f_{in}(p,\vec{x}-\frac{\vec{p}}{p^0}(t-t_0))+
\int_{t_0}^t B(\vec{x}-\frac{\vec{p}}{p^0}(t-t'),t') \,D(t_0,t') dt'\right]
\label{btesoln}
\end{equation}
where $D(t_2,t_1)$ is given by,
\begin{equation}
D(t_2,t_1)=\exp\left[-\int_{t_1}^{t_2} dt^\prime A(p,\,\vec{x}-\frac{\vec{p}}{p^0}(t'-t_0),\,t')\right]
\end{equation}
with
\begin{equation}
A(p,\vec{x},t)=\frac{p^0u_0(x)-\vec{p}\cdot\vec{u}(x)}{p^0 \tau_R(x)}
\end{equation}
and
\begin{equation}
B(\vec{x},t)=-\left( \frac{\partial}{\partial t}+
\frac{\vec{p}}{p^0}\cdot \frac{\partial}{\partial \vec{x} }\right)f_0(x,p).
\end{equation}
In equilibrium, $f_0$ for (Boson, say) is given by:
\begin{equation}
f_0(x,p)=\frac{1}{e^{\beta(x)(u^\mu p_\mu)- 1}}
\end{equation}
for which the expression for $B$ reduces to:
\begin{equation}
B(\vec{x},t)=-f_{eq}(1+f_{eq})\frac{p^{\mu}}{p^{0}}\partial_{\mu}\left[\beta(x)u^\mu p_\mu\right]
\end{equation}
where $\beta={1}/{T(x)}$,
$u^\mu(x)=(\gamma,\gamma\,\vec{v})$ is the flow velocity of the fluid  and the
Lorentz factor is given by,
$\gamma(x)=u^0(x)=(1-v(x)^2)^{-1/2}$. The space time dependence of $T$ and $\vec{v}$
are determined by the solution of the hydrodynamic equations.

Once $\delta f$ is known, perturbations in various 
thermodynamic quantities {\it e.g.} in 
energy density ($\epsilon$), entropy density ($s$), temperature etc can be 
obtained as follows. Any deviation  from the equilibrium value 
in the thermodynamic  quantities may be incorporated through the deviation in the 
distribution function, as discussed the  distribution function may be written as:
\begin{equation}
f(\vec{x},\vec{p},t )=f_0(p)\{1+\Psi(\vec{x},\vec{p},t)\},
\label{eq1}
\end{equation} 
The energy-momentum tensor ($T^{\mu\nu}$) of the system may be expressed in terms of $f$ as follows:
\begin{equation}
T^{\mu\nu}=\int d^3p\,\frac{p^{\mu}p^{\nu}}{p^0}f^{(0)}(p)\{1+\Psi(\vec{x},\vec{p},t)\}.
\label{eq2}
\end{equation}
where $T^{\mu\nu}$ can be split into equilibrium part ($\overline{T}^{\mu\nu}$) and a deviation,
$\Delta T^{\mu\nu}$ from the equilibrium value {\it i.e.} 
$T^{\mu\nu}=\overline{T}^{\mu\nu}+\Delta T^{\mu\nu}$. The tensor $T^{\mu\nu}$ can be 
decomposed into various components in terms of thermodynamic quantities as:
\begin{equation}
\begin{aligned}
{T}^{0}_{0}(x_i,t) &=-\{\epsilon+ \delta \epsilon(x_i,t)\},\\ 
{T}^{0}_{i}(x_i,t) &=-{T}^{i}_{0}=(\epsilon+P)u_i,\\
{T}^{i}_{j}(x_i,t)  &=\{P+\delta P(x_i,t) \}\delta^{i}_{j}+\Sigma^i_j(x_i,t),\\  
\Delta T^i_i=0,  
\end{aligned}
\label{eq5}
\end{equation}
where $P$ is the average pressure, $u_i$ is the $i^{th}$ (spatial) component of the flow velocity and
$\Sigma^i_j$ is the stress tensor which contains the shear viscous coefficient (see~\cite{sarwar} for details). 
The evolution of the flow velocity, energy density, pressure, temperature etc 
can be obtained from the solution of hydrodynamic equations.
The interaction of the expanding
background (hydrodynamics) with the perturbations ($\delta f$) at each space-time point is 
enforced through temperature (appearing through $\tau_R$) and flow velocity which 
are obtained from the solution of the relativistic hydrodynamic equations. Therefore,
the fluctuations represent interaction between equilibrium (hydrodynamics) and non-equilibrium
(BTE) degrees of freedom. We assume that the effects of the out-of-equilibrium perturbation 
on the equilibriated background is negligibly small.

The formalism discussed above can be used to any system where the fluctuations
are evolving in an expanding background aided by:
(a) initial distribution $\delta f_{in}$ appearing in Eq.~\ref{btesoln}, (b) $\tau_R$ which 
is determined  by the interaction at the microscopic level,  (c) flow velocity
and temperature determined by the solutions of hydrodynamic equations which
needs inputs like initial energy density and velocity distributions as well as EoS
controlled by the interactions in the system under study. 

In the present work we will apply this formalism to the system formed in HICRE. Therefore,
we will use  QCD based calculations for estimating $\tau_R(T)$ as:
$\tau_R^{-1}(x)=1.1\alpha_s T$ 
performed in Ref.~\cite{thoma} in HTL (Hard Thermal Loop approximation). We
have used QCD equation of state (section II.B) for solving hydrodynamical equations.

The power spectra have been evaluated for the following two scenarios: (i) for the fluctuation in 
the initial energy density obtained in  OG and MCG models,
(ii) for fluctuations caused by perturbations in phase space distribution.
The latter one has been evolved through BTE in an expanding thermal QGP background as discussed.
To simulate different types of initial spatial anisotropy one may choose,
\begin{equation}
\delta f(p,\vec{x},t_0)=A_0\exp\left[-r(1+a_n \cos n\vartheta)\right]
\label{initialdist}
\end{equation}
where $A_0$ is a constant and $n$ can be taken as $n= 2,3,4,5,...$ to simulate different 
geometry for the initial anisotropy (see also~\cite{sorensen}). We have taken $A_0=1$ 
and set the perturbation centered around $r=0$ here. 
\section{The power spectrum}
We are now equipped to study the evolution of the power spectrum of the angular
distribution of particles which
originates due to:  (i) fluctuations in initial energy density profile
evaluated in  OG and MCG models by using the momentum distributions of particles 
at various surfaces of constant temperatures with the help of mathematical
expression given in~\cite{CF}:
\be
E\frac{d N_0}{d^3p}=\frac{g_i}{(2 \pi)^3}\int_{\Sigma} d\sigma_\mu p^\mu f_0(x,p)
\ee
where  $d\sigma_\mu$ is the surface element, $p^\mu$ is the 4-momenta of the particle and 
in Milne coordinate these are expressed as follows~\cite{hlle,jeonheinz}:
\begin{align*}
d\sigma _{\mu}&=(\tau_f\mbox{ }dx_f\mbox{ }dy_f\mbox{ }d\eta_f,-\tau_f\mbox{ }
d\tau _f\mbox{ }dy_f\mbox{ }d\eta_f,-\tau_f\mbox{ }d\tau _f\mbox{ }dx_f\mbox{ }d\eta_f,
-\tau_f\mbox{ }d\tau_f \mbox{ }dx_f\mbox{ }dy_f)\\
&=\left( 1,-\frac{\partial \tau _f}{\partial x_f},-\frac{\partial \tau _f}{\partial y_f},
-\frac{\partial \tau _f}{\partial \eta_f}\right)\tau_f \mbox{ }dx_f\mbox{ }dy_f\mbox{ }d\eta_f \\
p^{\mu}&=(m_T\mbox{cosh}(y-\eta_f),p_x,p_y,m_T\mbox{sinh}(y-\eta_f)/\tau_f)
\label{Xpu}
\end{align*}
where $\eta_f$ is the fluid rapidity, $\tau_f$ is the proper time,
$x_f$ and $y_f$ are transverse coordinate for the fluid.
$p_x$, $p_y$ are the fluid momenta in Cartesian coordinate, $y$ is the
particle rapidity and $m_T=\sqrt{p_T^2+m^2}$ is the transverse mass
of the particle. In the above equation subscript $f$ stands for fluid. 
Therefore,
$$p^{\mu}d\sigma _{\mu}=\left[ m_T\mbox{cosh}(y-\eta_f)-p_x\frac{\partial \tau _f}{\partial x}-p_y\frac{\partial 
\tau _f}{\partial y}-\frac{m_T\mbox{sinh}(y-\eta_f)}{\tau _f}\frac{\partial \tau _f}
{\partial \eta_f}\right]\tau_f \mbox{ }dx_f\mbox{ }dy_f\mbox{ }d\eta_f$$ 
For massless particles, $y$ is given by
$$y=-\text{ln}\left[ \text{tan}\left( \frac{\theta}{2}\right)\right]$$
which is same as the pseudo-rapidity ($\eta$) of the particle.

(ii)Similarly the fluctuation in the particle distribution due to the perturbation, $\delta f$ 
in phase space distribution obtained  within the 
framework of relaxation time approximation by solving BTE in an expanding QGP background.
$\delta f$ can be used to estimate perturbations thermodynamic quantities as mentioned earlier.
We use $f=f_0+\delta f$ to estimate the $p_T$  distribution of particles away from equilibrium as: 
\be
E\frac{dN}{d^3p}=\frac{g_i}{(2 \pi)^3}\int_{\Sigma} d\sigma_\mu p^\mu [f_0+\delta f(x,p)]
\ee
Now the quantities, $EdN_0/d^3p$   or $EdN/d^3p$ 
can be expanded in Laplace series in terms of spherical harmonics, 
$Y_{lm}(\theta,\phi $) at a given 
transverse momentum ($p_T$) and  temperature.
We identify pseudo-rapidity, $\eta$ as the polar angle through the relation, $\eta=-ln\{tan(\theta/2)\}$.

The power spectrum of the fluctuations in the transverse 
momentum ($p_T$) distribution of particles can be estimated at
surfaces of constant temperatures to understand its evolution as the system 
cooled down with the progression of time. 
The power spectrum of $EdN/d^3p$ has been estimated as follows:
\be
E\frac{dN}{d^3p}=\bar{N}+\sum_{l=1}^{\infty}\sum_{m=-l}^{l}a_{lm}(p_T,T) Y_{lm}(\theta,\phi) 
\label{ylm}
\ee
where 
\begin{equation}
\bar{N}=\frac{1}{4\pi}\int d\Omega\frac{dN}{d^2p_Tdy}
\end{equation} 
the coefficients, $a_{lm}$'s are determined  as follows:
\be
a_{lm}(p_T,T)=\int d\Omega Y_{lm}^{\ast} E\frac{dN}{d^3p}
\label{eqalm}
\ee
For determining power spectrum without perturbation we replace
$EdN/d^3p$ by $EdN_0/d^3p$ in Eq.~\ref{eqalm}.
The terms in Eq.~\ref{ylm} with different $l$ 
corresponds to different angular scales: terms with larger $l$ will have   
smaller angular resolution,  
$\theta_l=\pi/l$ ~\cite{cmbr} determines the 
value of the maximum $l$ i.e. $l_{max}$. For heavy ion experiments at RHIC and
LHC the resolution in pseudo-rapidity will govern the value of $l_{max}$.

Using standard techniques and properties of spherical harmonics, 
the angular power spectrum ($C_l$) of $EdN/d^3p$ can be written as:
\be 
C_l(p_T,T)=\frac{1}{2l+1}\,\sum_m |a_{lm}|^2 
\ee
indicating the  distribution of power of fluctuations among different angular scales
determined by $l$. 
In CMBR fluctuation, $C_l$ for the temperature fluctuation, $\Delta T(\theta,\phi)/T$
has been calculated theoretically and compared with experimental data which has enhanced our
understanding on the matter content of the universe.
The power spectrum, $C_l$'s are related to the various flow harmonics~\cite{jadams,Ollitra,uheinz,arnold,HC}
as shown in appendix B.
In analogy with CMBR, the solution of BTE  can be used to estimate the 
fluctuations in temperature $(\Delta T/T$) as:
\begin{equation}
\frac{\Delta T}{T}=-\left(\frac{\partial lnf_0}{\partial lnp}\right)^{-1}\Psi
\label{delT}
\end{equation}

\begin{figure}
\centerline{\includegraphics[height=100mm, width=100mm]{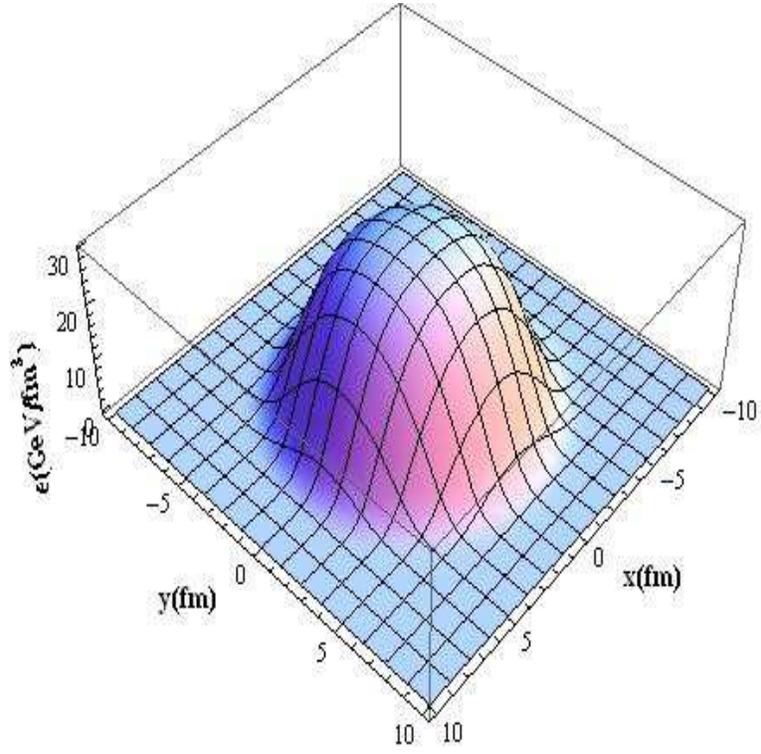}}
\caption{Initial energy density profile in the transverse plane at 
space time rapidity = 0 in the OG model for Au+Au central collision.
}
\label{fig1}
\end{figure}

\section{Results}
We have solved the (3+1) dimensional relativistic ideal hydrodynamic
equations with initial conditions  (due to OG and MCG models)
and EoS described above to study the evolving  QGP.  
The Boltzmann transport equation has been solved in relaxation
time approximation in this expanding background to study the effects of phase space 
perturbation. The  solution of BTE can be used
to estimate the power spectrum due to phase space perturbations in various physical quantities.
We present the results for the following two sets of conditions: 

(i) The power spectrum of $EdN_0/d^3p$ due to fluctuations in initial energy density.
We estimate the power spectrum  
of the $p_T$ distributions of particles ($dN_0/d^2p_Tdy$)
for OG and MCG model initial conditions at surfaces of constant
temperatures, say $T_S$, defined as $T(x,y,\tau,\eta_s)=T_S$ (where $\eta_s$ 
is space-time rapidity) which implies, $\tau=\tau(x,y)$ on the surface. 
We will evaluate the power spectrum at $T=T_S=350$ MeV which is
close to the initial temperature, near the transition temperature ($T_c$),
$T_S=T_c=170$ MeV and at some intermediary temperature, $T=T_S=250$ MeV
to understand how the power spectrum evolve from the initial to the transition point. 

(ii) The power spectrum of $EdN/d^3p$  which contains the perturbations has also been estimated. The perturbation 
in phase space has been obtained from the solution of BTE.

First we consider (i):
in Fig.~\ref{fig1} the  initial energy density profile due to OG model is displayed for
central Au + Au collision at  $\sqrt{s_{NN}}=200$ GeV. 
The thermalization time has been taken  
as $\tau_i=0.6$ fm/c.  Other parameters regarding 
the initial condition are displayed in table 1 in section II.A.
The profile evaluated at zero space-time rapidity has isotropic symmetry with sharp fall near the boundaries.

\begin{figure}
\centerline{\includegraphics[height=100mm, width=100mm]{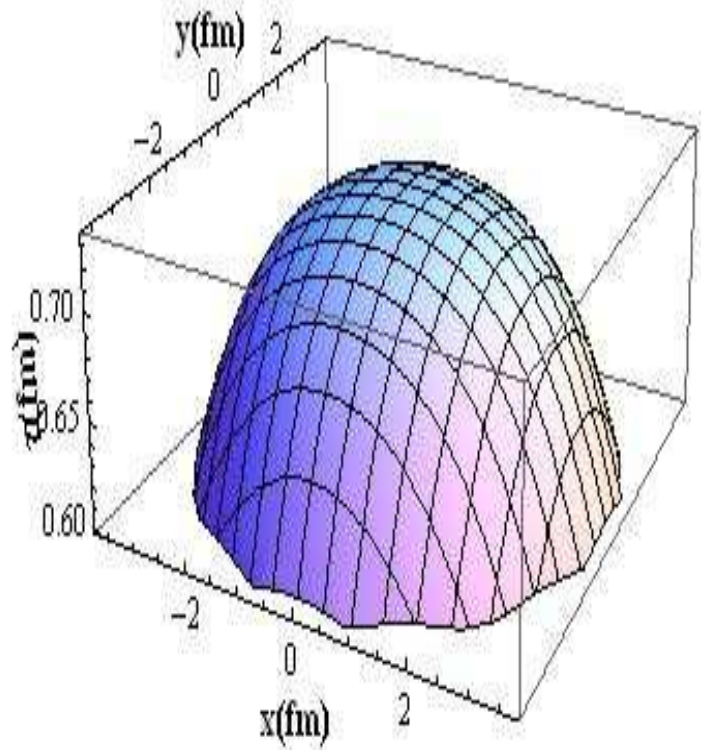}}
\caption{The constant temperature, $T=350$ MeV surface in the time-transverse plane.
}
\label{fig1a_1}
\end{figure}

\begin{figure}
\centerline{\includegraphics[height=100mm, width=100mm]{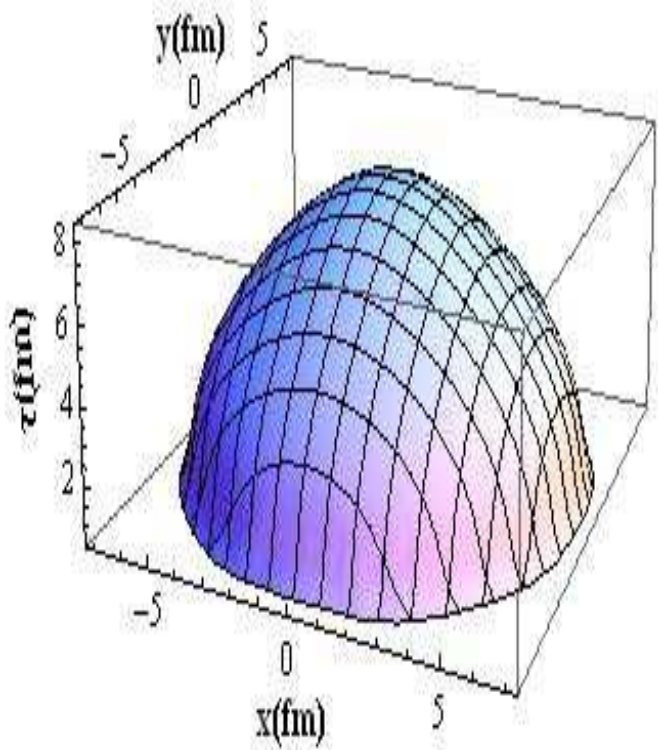}}
\caption{The constant temperature surface in the time-transverse plane
for $170$ MeV. 
}
\label{fig1a_2}
\end{figure}

In Figs.~\ref{fig1a_1} and ~\ref{fig1a_2}, the surfaces of constant temperatures evolved hydrodynamically 
for initial energy density (shown in Fig.~\ref{fig1}) have been depicted. 
Fig.~\ref{fig1a_1} (Fig.~\ref{fig1a_2}) shows the result for $T=350$ (170) MeV. 
We observe no qualitative change in the shape of the surfaces except at lower 
temperature the space-time size of the surface becomes larger. It is important to note
that the energy density profile and
consequently the constant temperature surfaces are smooth - not showing any
distinct fluctuations because of the  absence of fluctuation in the initial energy 
density profile in OG model.


\begin{figure}
\centerline{\includegraphics[height=100mm, width=100mm]{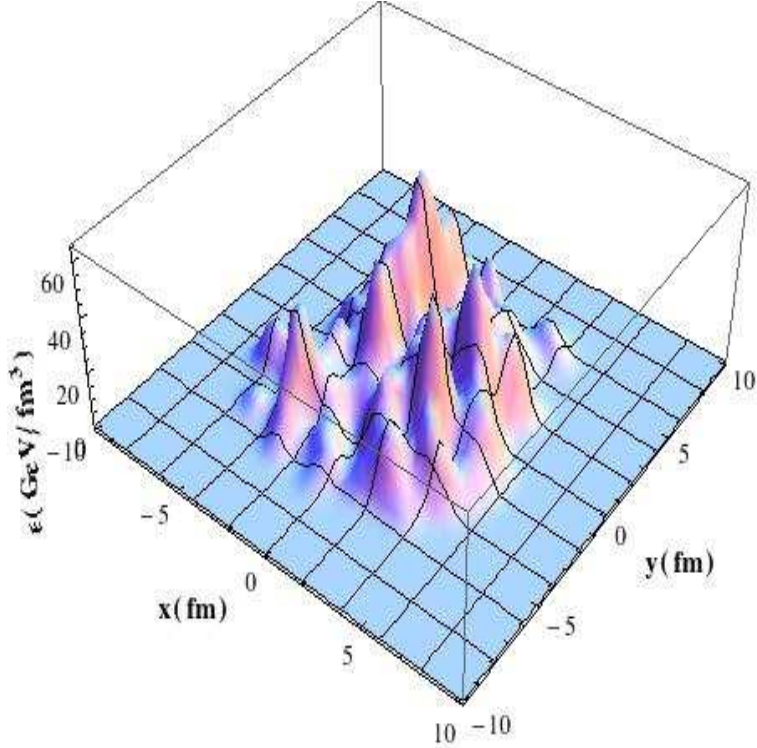}}
\caption{Same as Fig.~\ref{fig1} with MCG initial condition.
}
\label{fig2}
\end{figure}

Similar to the OG initial condition we plot the initial energy density profile
evaluated in MCG model for $0-5\%$ centrality collision of  Au + Au at 
$\sqrt{s_{NN}}=200$ GeV in Fig.~\ref{fig2} with $\tau_0=0.6$ fm.  
We observe lumpiness of complicated nature in the initial energy density
profile at various position in the transverse plane due to the collisions
of nucleons with  fluctuating positions in the beam nuclei.

\begin{figure}
\centerline{\includegraphics[height=100mm, width=100mm]{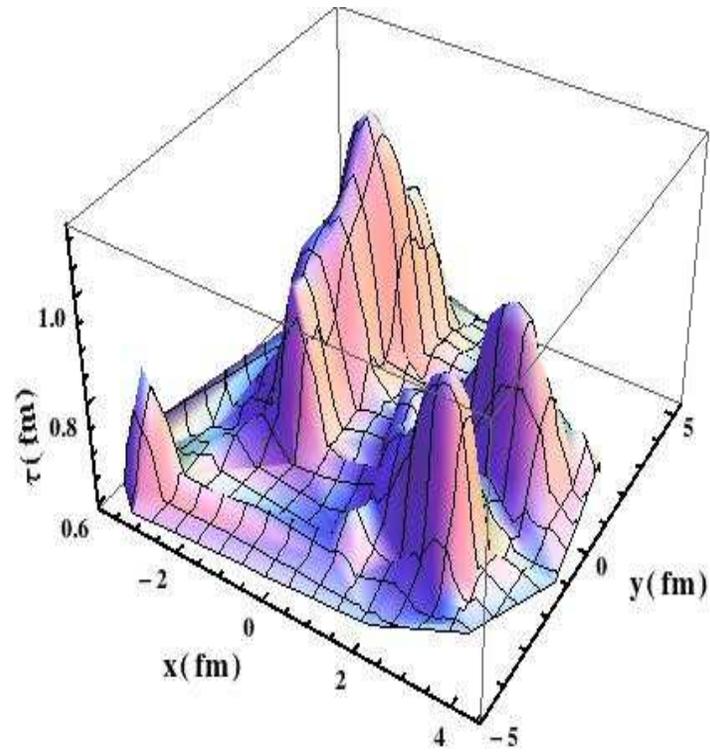}}
\caption{The constant temperature surface at $T=350$ MeV for MCG initial condition.
}
\label{fig2b_1}
\end{figure}

\begin{figure}
\centerline{\includegraphics[height=100mm, width=100mm]{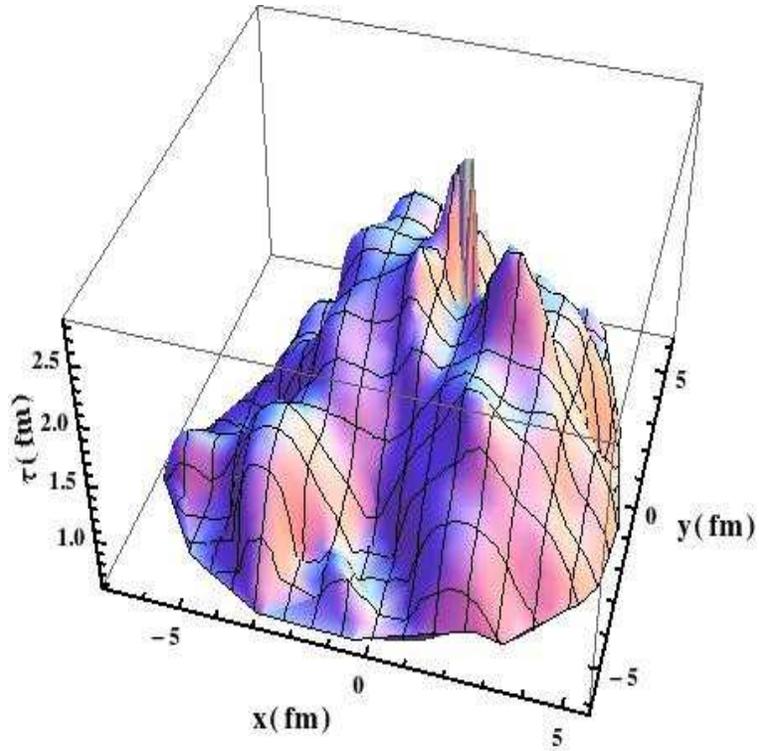}}
\caption{Same as Fig.~\ref{fig2b_1} for MCG initial condition at $T=250$ MeV.
}
\label{fig2b_1a}
\end{figure}

\begin{figure}
\centerline{\includegraphics[height=100mm, width=100mm]{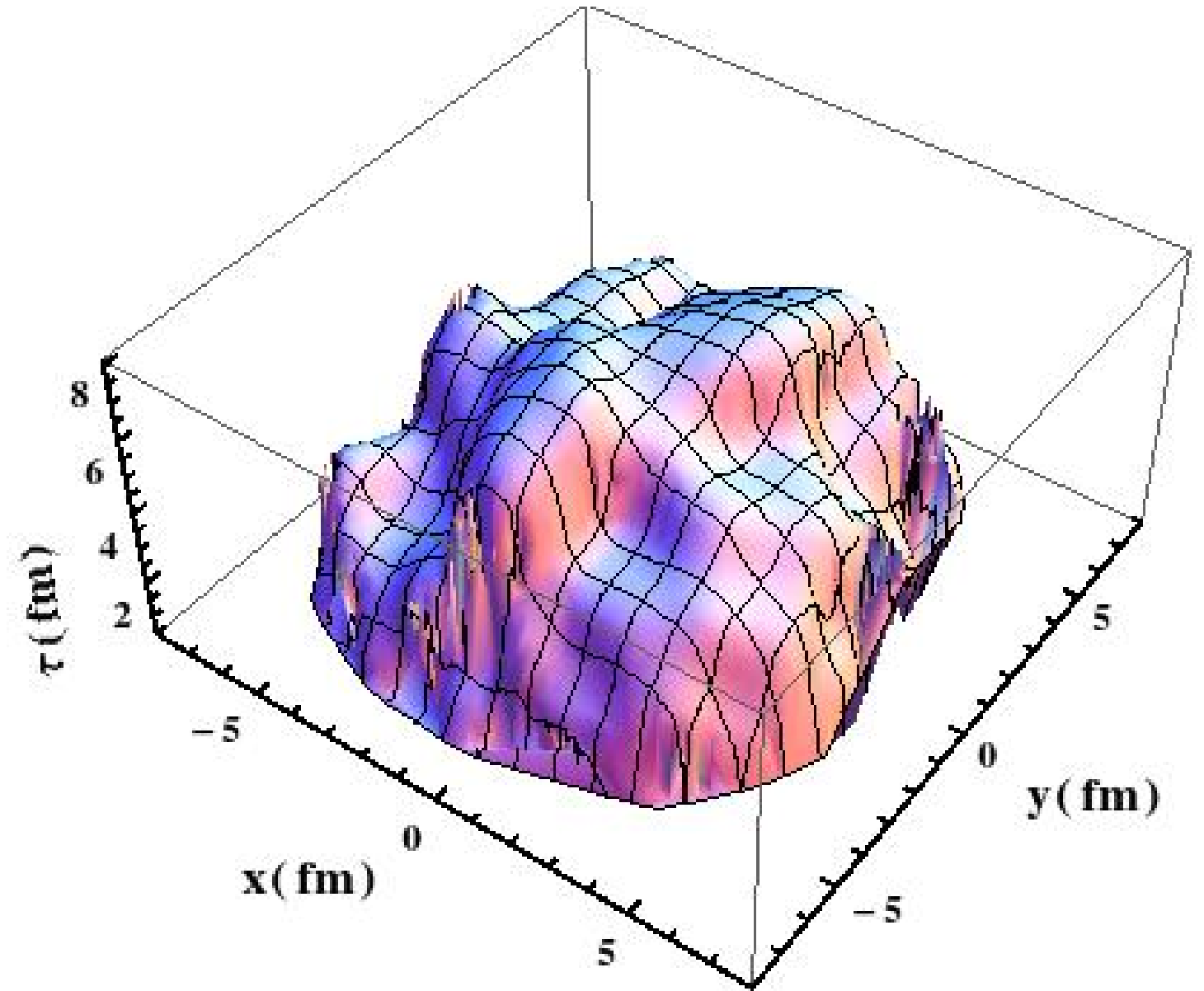}}
\caption{Same as Fig.~\ref{fig2b_1} at $T=170$ MeV.
}
\label{fig2b_2}
\end{figure}

The constant temperature surface at $T=350 (250)$ MeV is
displayed in $(\tau, x, y)$ coordinate in Fig.~\ref{fig2b_1}  (Fig.~\ref{fig2b_1a})
for MCG initial condition. It is observed that the initial energy density has significant
fluctuations in space time coordinate. These inhomogeneities will 
create pressure imbalance with the neighbouring zones  - higher density
domains will exert larger pressure and hence will expand faster to
smoothen the inhomogeneities. As a result the inhomogeneity  
will  reduce and their distributions will 
change. We observe that the size of the surface 
in space-time coordinate  has increased at lower temperature. 
The large fluctuations at the surface corresponding to $T=350$ MeV 
resulting from the inhomogeneities in the initial energy density profile have 
reduced in magnitude  at the $T=250$ MeV surface as the system evolve 
hydrodynamically.  The domains of high energy densities (Fig.~\ref{fig2}) 
will take longer time to reach a given temperature as can be seen
from Fig.~\ref{fig2b_1} where for certain domains $\tau$ is larger 
compared to others. 
We observe that the differences
in the values of $\tau(x,y)$ at various points in $x-y$ plane is smaller
in the $T=170$ MeV surface than at $T=350$ or 250 MeV surfaces. Indicating
that the system is approaching toward a homogeneous one in
coordinate space and through expansion this inhomogeneities
get transferred  to momentum space.
However, the magnitude of fluctuations have reduced in real space
at lower temperature, $T=170$ MeV (Fig.~\ref{fig2b_2}). 

\begin{figure}
\centerline{\includegraphics[height=120mm,width=90mm,angle=-90]{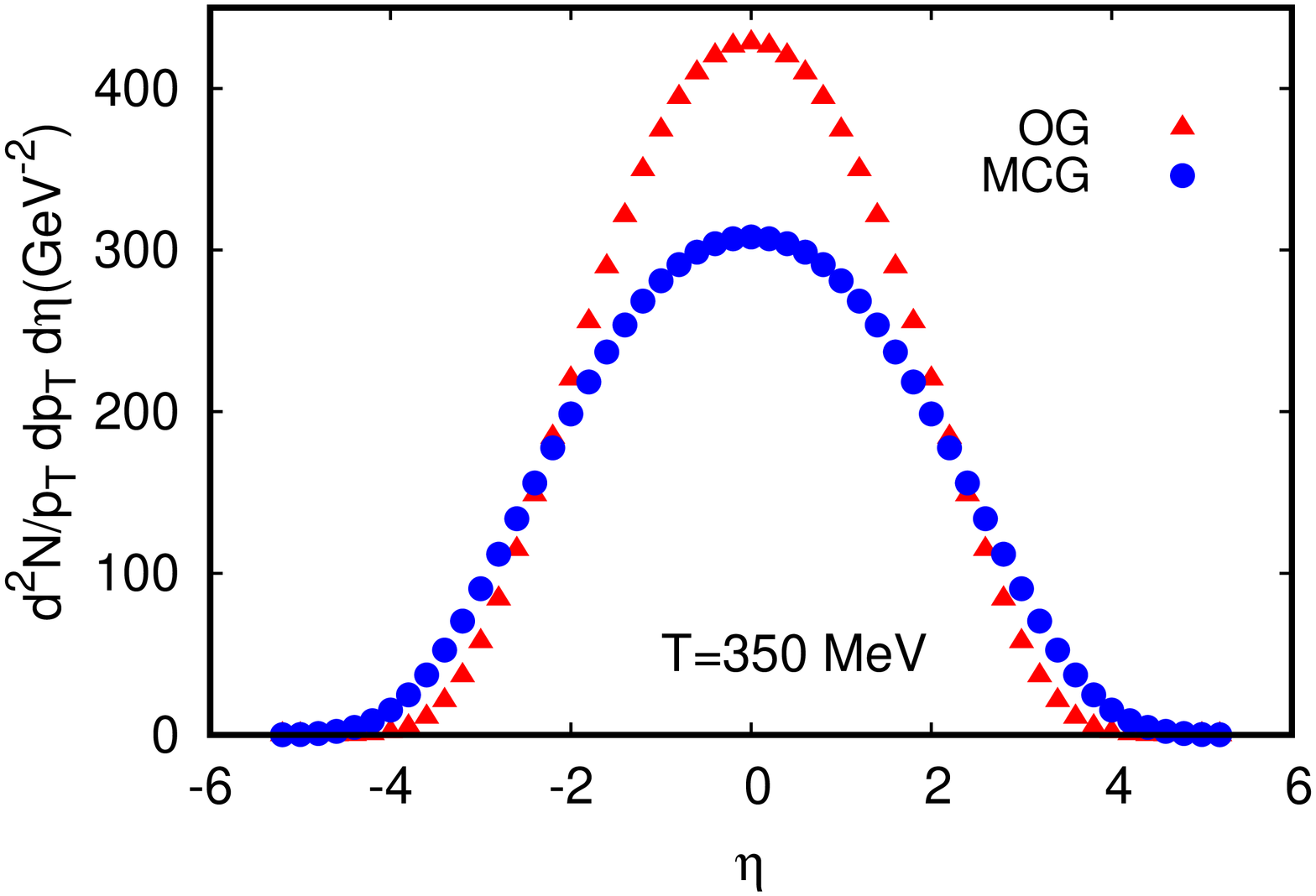}}
\centerline{\includegraphics[height=120mm,width=90mm,angle=-90]{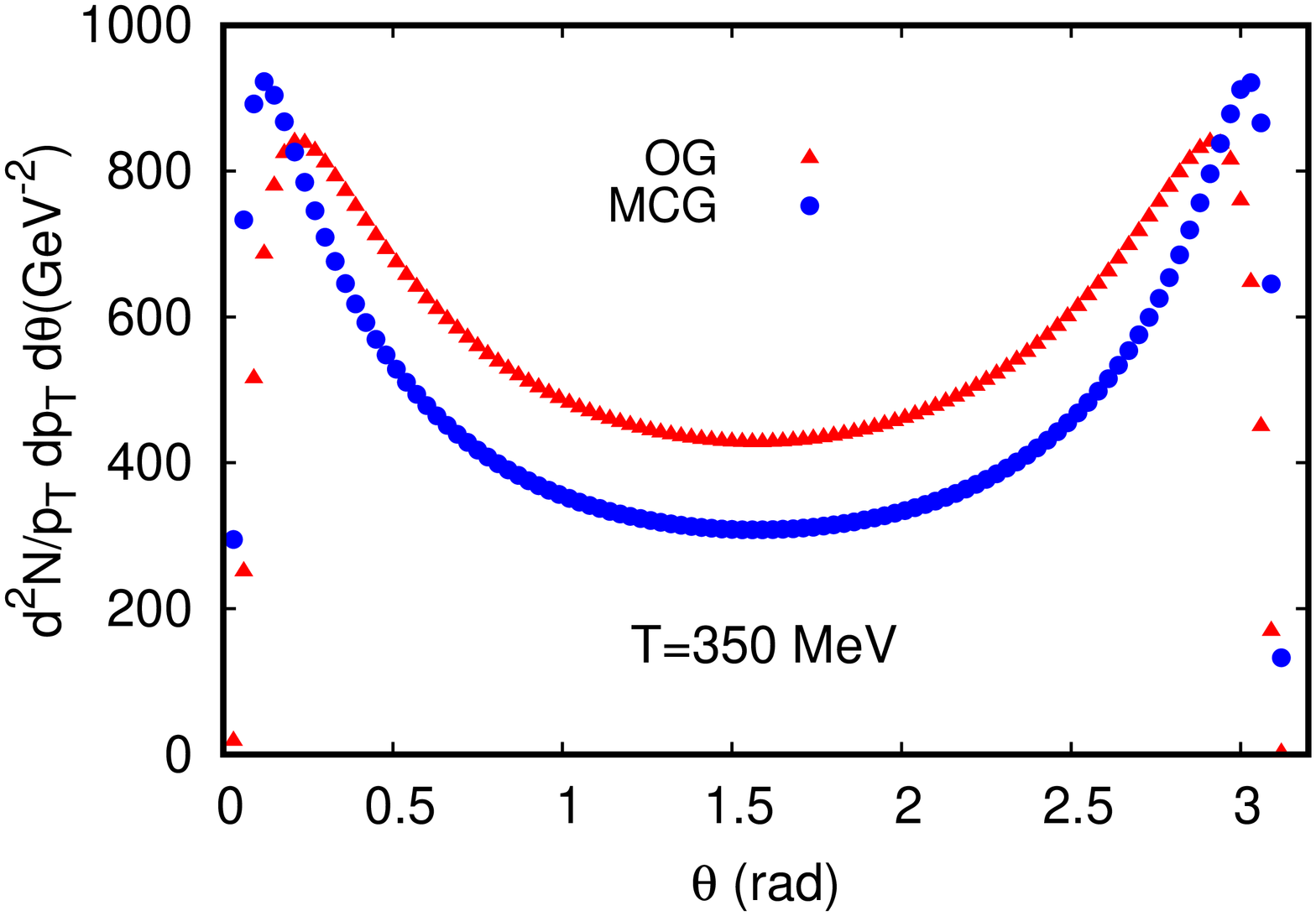}}
\caption{The pseudo-rapidity ($\eta$) and angular ($\theta$) 
distribution of particles at $p_T=0.6$ GeV with OG and MCG initial 
conditions at $T=350$ MeV.
}
\label{figetaT350}
\end{figure}

\begin{figure}
\centerline{\includegraphics[height=120mm,width=90mm,angle=-90]{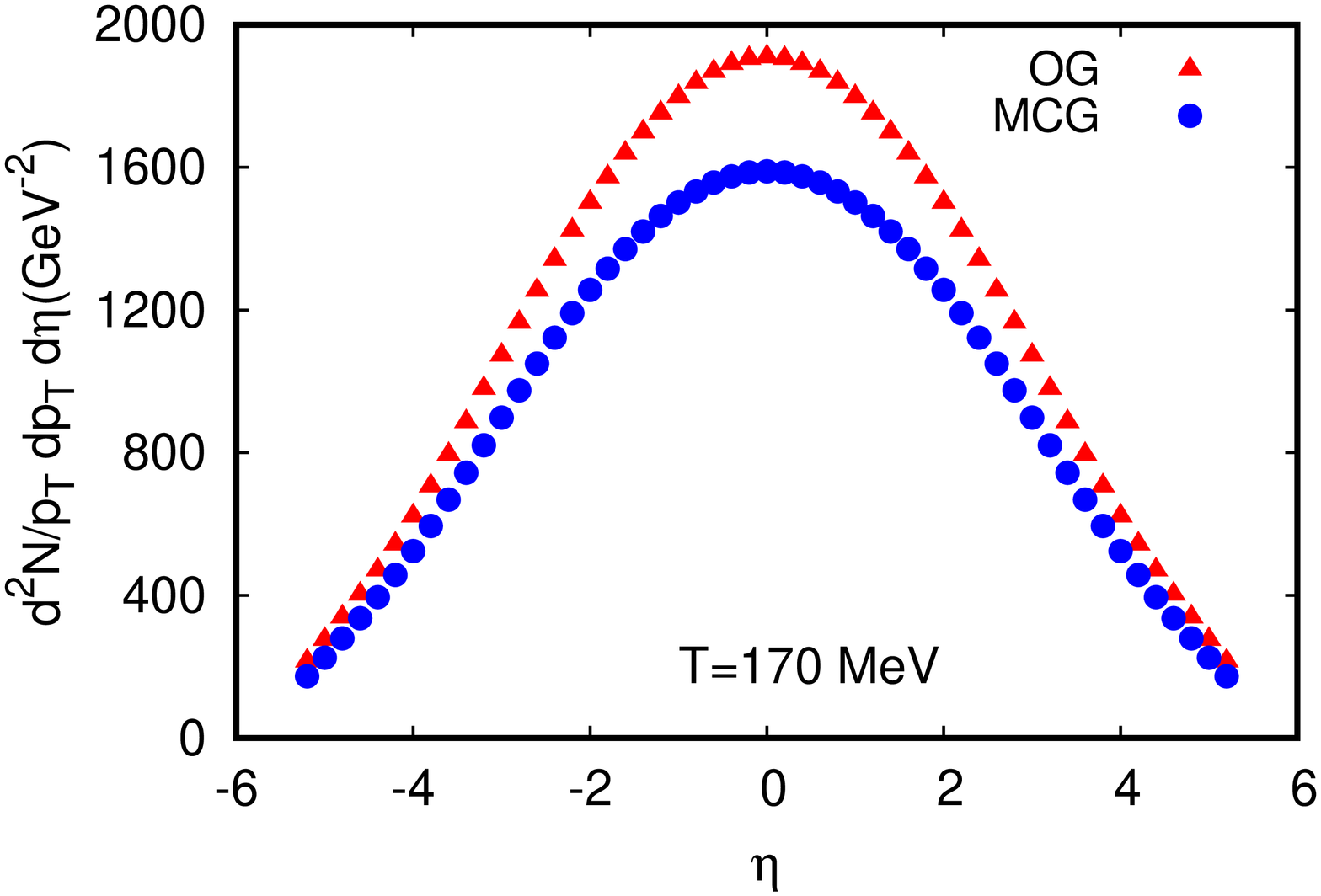}}
\centerline{\includegraphics[height=120mm,width=90mm,angle=-90]{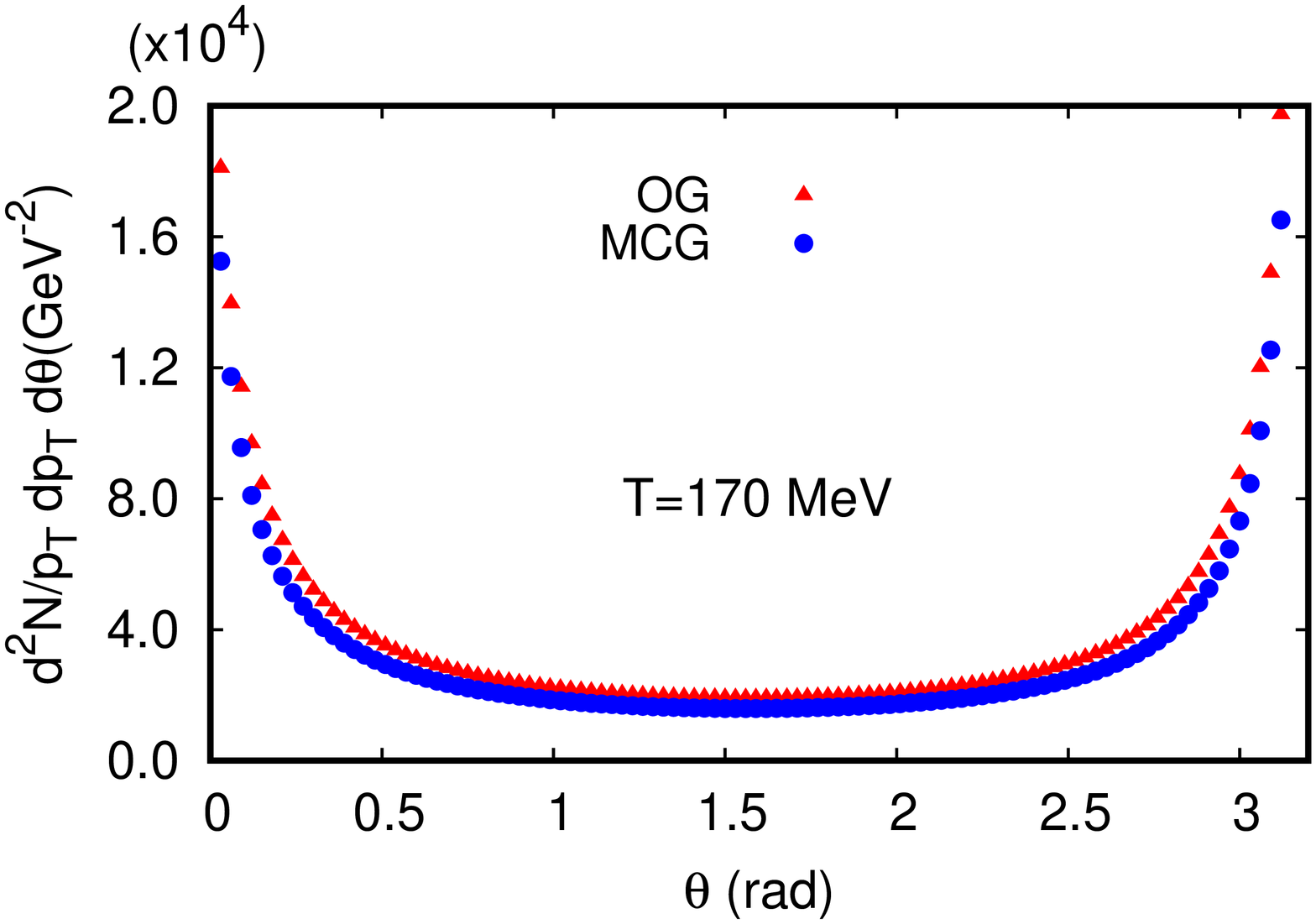}}
\caption{Same as Fig.~\ref{figetaT350} at $T=170$ MeV.
}
\label{figetaT170}
\end{figure}

In Fig.~\ref{figetaT350} the $\eta$ (upper panel) and $\theta$ (lower panel) distributions of particles 
have been displayed at the surface of constant temperatures, $T=350$ MeV. We find  that
the fluctuations in differential particle numbers is larger at MCG than OG model, which is
clearly visible both in the $\eta$ and $\theta$ distributions. However, with progress in
time or with the reduction of temperature at $T=170$ MeV (Fig.~\ref{figetaT170}) the total number of particles 
increase and more particles appear at larger $\eta$ enhancing the width of the distributions.
The $\theta$ distribution shows a plateau over a larger domain of $\theta$  

Next we would like to study - how the power spectrum of the fluctuations
caused by initial energy density profiles 
evolves. We will study the power spectrum of particle spectra,
$EdN_0/d^3p$  at the surfaces of constant temperatures 
at $T=350, 250$ and 170 MeV. The angular distribution of particles 
at constant $p_T$ has been analyzed by decomposing it in terms
of spherical harmonics as discussed in section IV.

\begin{figure}
\centerline{\includegraphics[height=120mm,width=90mm,angle=-90]{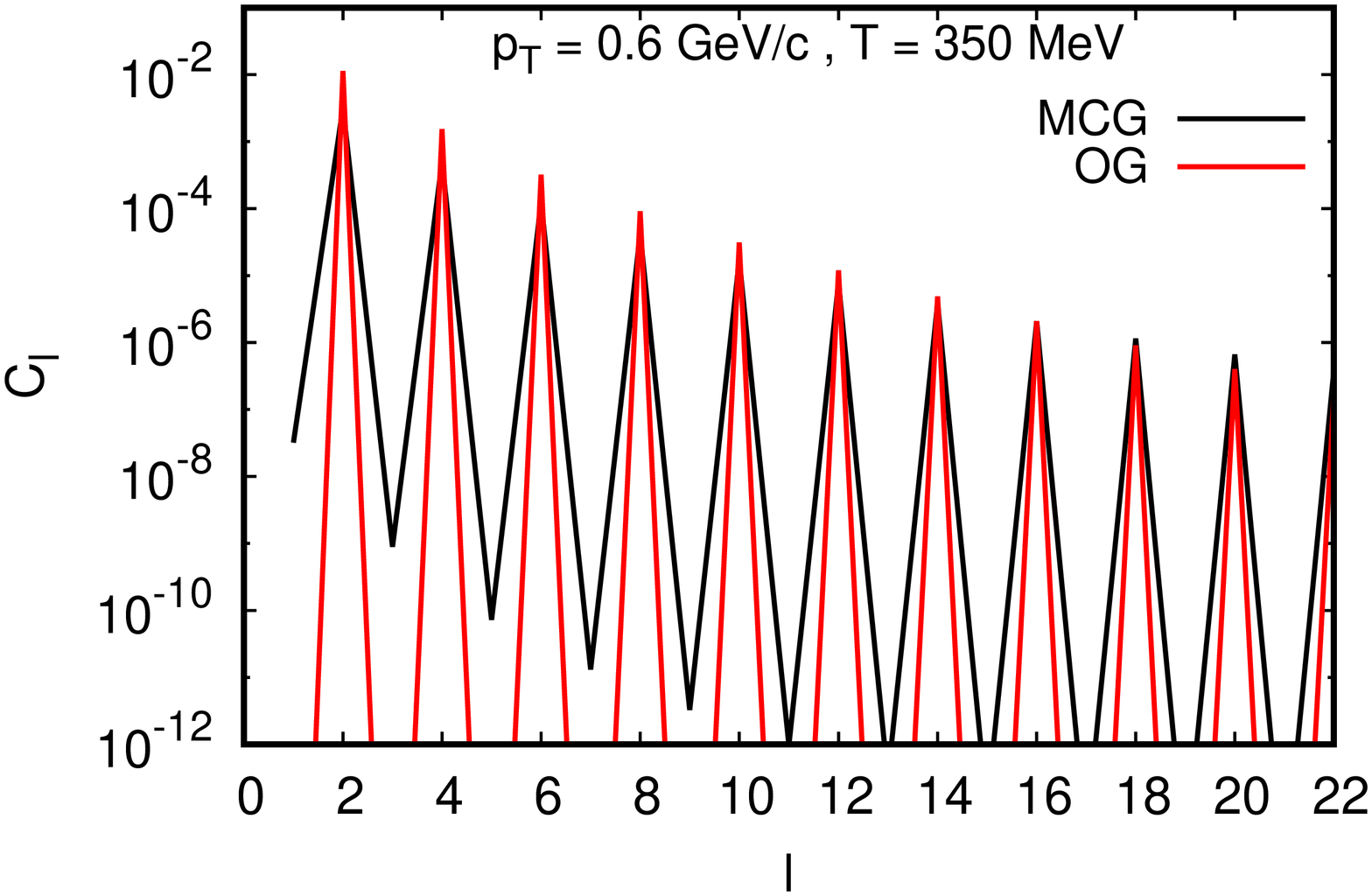}}
\caption{The power spectrum, $C_l$ deduced from $dN/d^2p_Tdy$ at 
$p_T=0.6$ GeV for both the OG and MCG initial conditions
analyzed at the surface of $T=350$ MeV. 
}
\label{fig3}
\end{figure}

The power spectrum of the distribution,
$C_l$ has been plotted in Fig.~\ref{fig3} for the angular
distribution of  the spectra at $T=350$ MeV for the OG initial condition.
We clearly find that the power spectrum corresponding to the odd $l$'s 
are negligibly small because the distribution is an even function of $\theta$. 


\begin{figure}
\centerline{\includegraphics[height=120mm,width=90mm,angle=-90]{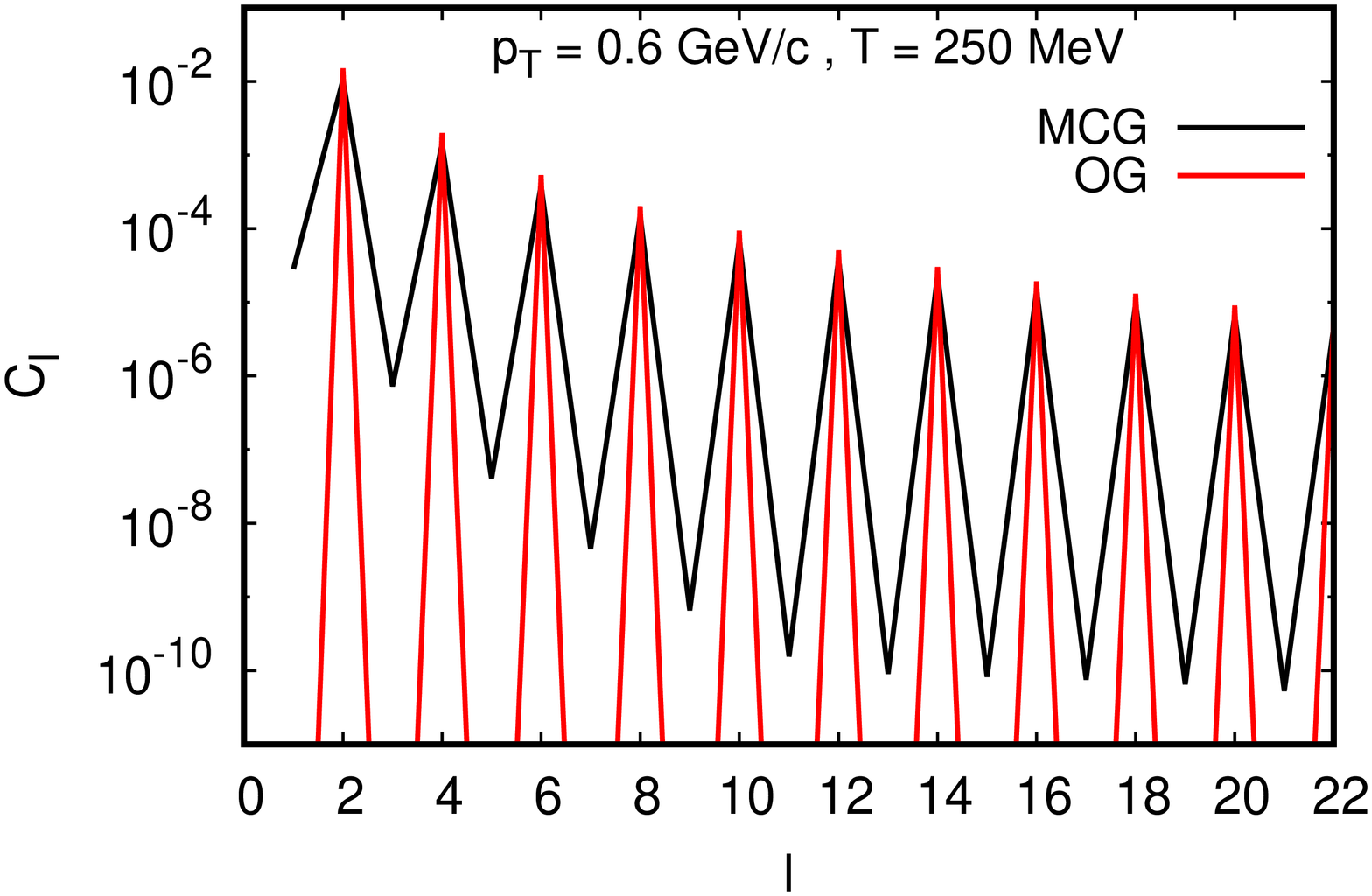}}
\caption{Same as Fig.~\ref{fig3} at $T=250$ MeV.
}
\label{fig4}
\end{figure}

\begin{figure}
\centerline{\includegraphics[height=120mm,width=90mm,angle=-90]{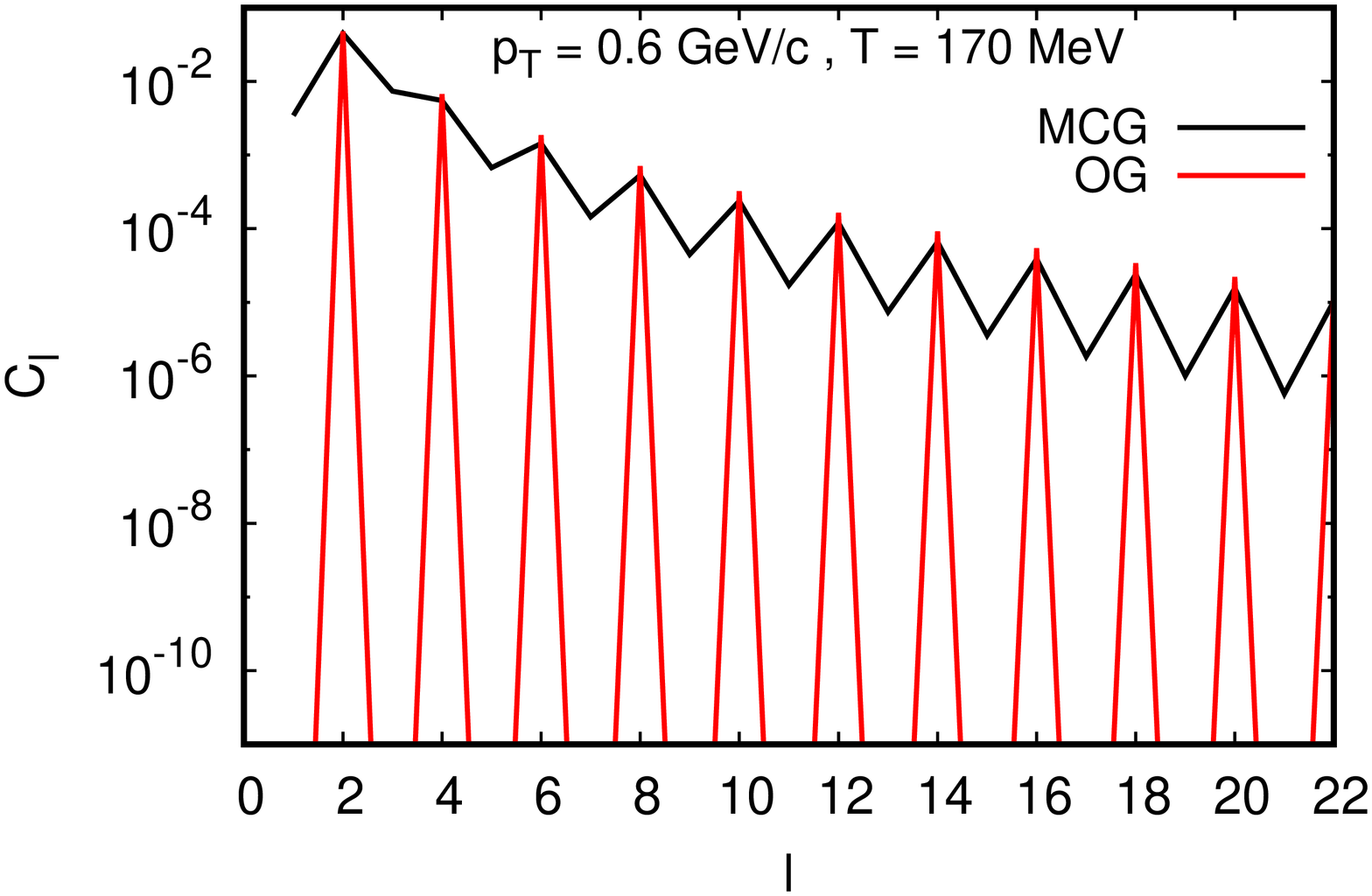}}
\caption{Same as Fig.~\ref{fig3} at $T=170$ MeV.
}
\label{fig5}
\end{figure}

In Figs.~\ref{fig4} and ~\ref{fig5} the power spectrum for the 
angular distribution
at the the surface of $T=250$ and 170 MeV respectively have been depicted. 
We observe that there is no significant change in the 
power spectrum for the OG initial conditions at lower temperatures.
With time, the spatial  inhomogeneities in $x-y$ plane (which are translated into momentum space
due to force caused by pressure gradient)
of the system gets reduced
as the system favours to erase out any pressure imbalance,  
however, for systems without  fluctuation as in OG case, does not show
much change. In case of OG initial conditions    
the system is symmetric, (Fig.~\ref{fig1})
therefore, with the evolution from higher to lower temperatures 
(Fig.~\ref{fig1a_1} and \ref{fig1a_2})
there is no significant change in the power spectrum. 
\begin{figure}
\centerline{\includegraphics[height=100mm, width=100mm]{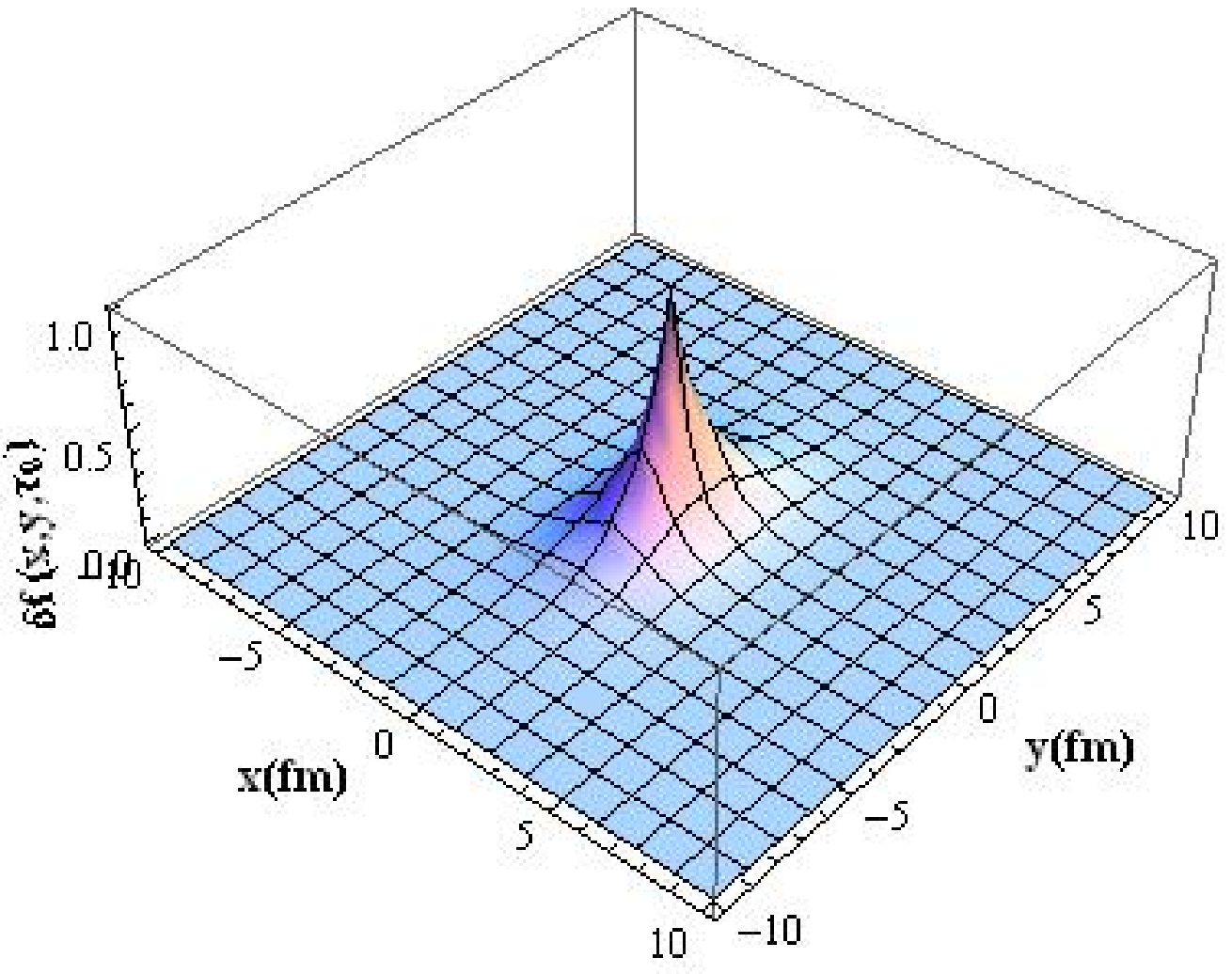}}
\caption{The initial perturbation $\delta f$ given in Eq.~\ref{initialdist}
with $n=2$.
}
\label{fig6}
\end{figure}

However, for MCG initial condition, the system
is highly inhomogeneous (Fig.~\ref{fig2}). The pressure gradients
caused by the inhomogeneity  acts in favour of reducing it
during the course of expansion from higher to 
lower temperature. Therefore, it is clearly seen that the power spectrum
of the system appears to be different at $T=250$ MeV (Fig.~\ref{fig4}) as the contribution from 
odd $l$ are enhanced compared its value at $T=350$ MeV (Fig.~\ref{fig3}). 
We also note that $C_l$'s increases with lowering of
temperatures both for OG and MCG initial conditions (see later).
It is well known that smaller (larger) $l$'s resolve larger (smaller) angular
anisotropy. The value of $l$ sets the angular scale, $\theta_l=\pi/l$. 
The most interesting aspect is that at lower temperatures ($T=170$ MeV, Fig.~\ref{fig5}), the power
spectrum at odd $l$'s become comparable in magnitude to the values at even $l$'s.  
The enhancement of the odd $l$'s is a signature of the presence of inhomogeneities 
in the initial condition. 
The increase in $C_l$'s for large $l$ at lower temperature ($T=170$ MeV) indicates appearance 
of smaller angular fluctuations.

\begin{figure}
\centerline{\includegraphics[height=120mm,width=90mm,angle=-90]{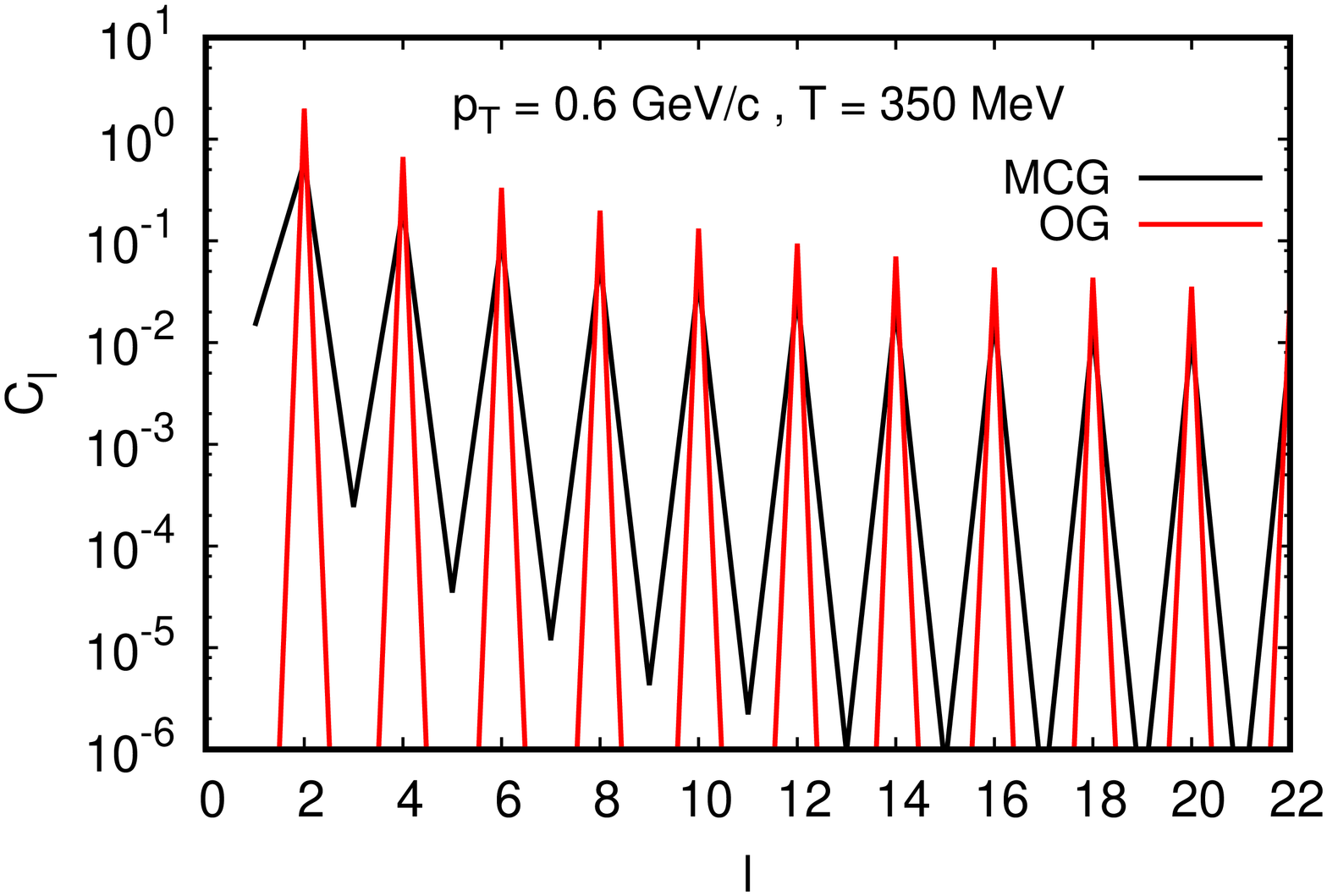}}
\caption{The power spectrum of the perturbation at $T = 350 \text{ MeV}$ for perturbation shown in Fig~\ref{fig5}. 
The red (black) line shows results for OG (MCG) initial conditions for $p_T = 0.6 \text{ GeV}/c$}
\label{fig7}
\end{figure}
\begin{figure}
\centerline{\includegraphics[height=120mm,width=90mm,angle=-90]{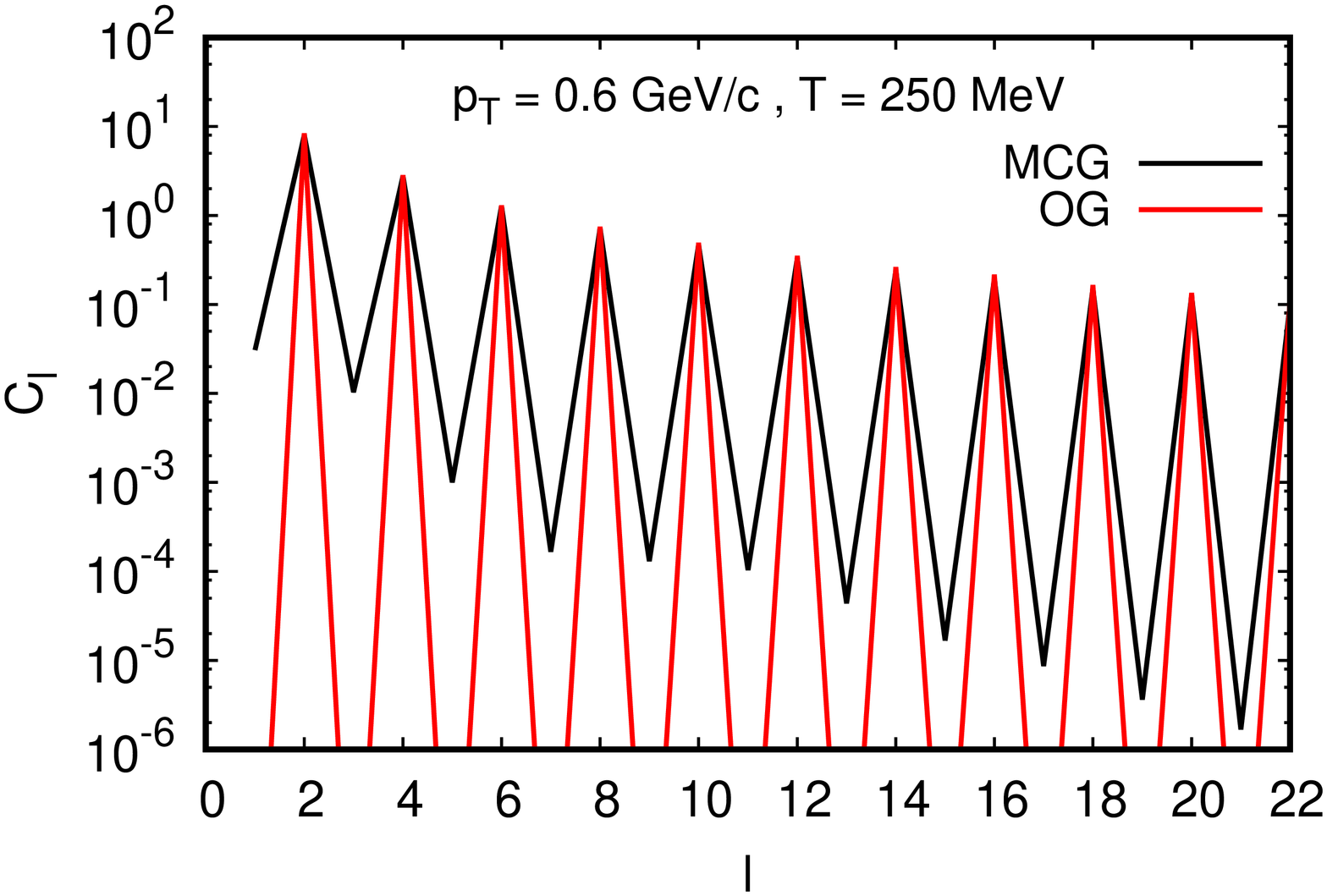}}
\caption{Same as Fig~\ref{fig7} at $T= 250\text{ MeV}$}
\label{fig8}
\end{figure}

(ii) So far we have discussed the evolution of power spectrum created 
in the initial collision dynamics. However, fluctuations  represented by
$\delta f$ may be
caused by other sources also, {\it e.g.}  propagation of
jets through the medium may create  such fluctuations.
Therefore, next we make some case studies
on the propagation of fluctuations by introducing
perturbations $\delta f(\vec{p},\vec{x},t)$ in phase space distribution
by solving BTE in expanding background. In Fig.~\ref{fig6} the initial perturbation $\delta f$
(Eq.~\ref{initialdist} for $n=2$) has been displayed. 
\begin{figure}
\centerline{\includegraphics[height=120mm,width=90mm,angle=-90]{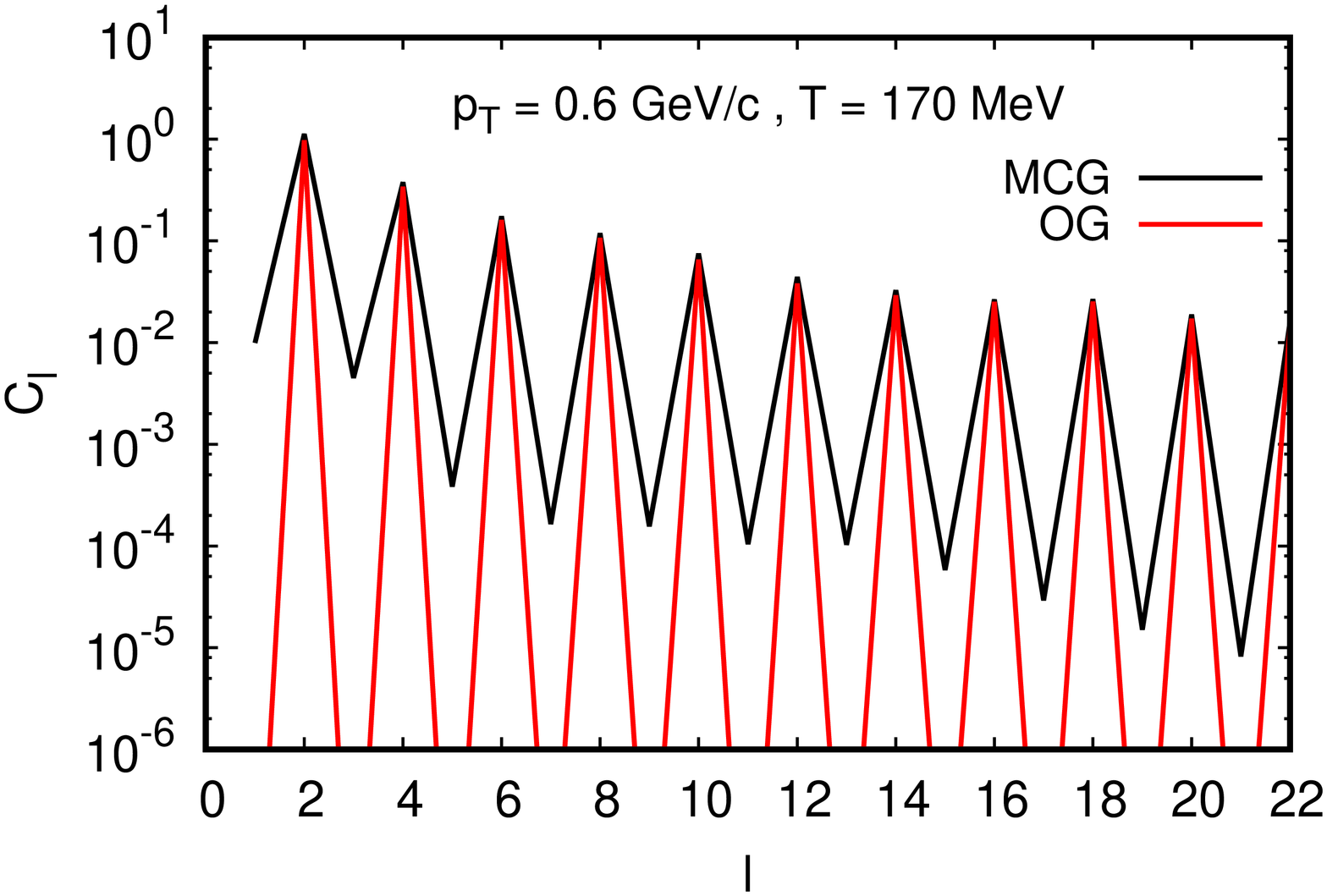}}
\caption{Same as Fig.~\ref{fig3} at $T=170$ MeV.
}
\label{fig9}
\end{figure}
We study the power spectrum of $E d\delta N/d^3p$ for the perturbation shown in 
Fig.~\ref{fig6}. $C_l$ at $T=350, 250$ and 170 MeV, for both
OG and MCG are depicted in Figs.~\ref{fig7}, ~\ref{fig8} and ~\ref{fig9} for $p_T=0.6$ GeV. 
We recall that there is negligible inhomogeneity in background for OG initial condition whereas 
the background inhomogeneity MCG initial condition is large. 
Therefore, this study gives us an opportunity to learn
how the local fluctuations (due to $\delta f$) evolve with the expanding background inhomogeneities.
We find that the spectrum with OG initial condition remains largely unaltered. However,
the amplitude of the spectrum with MCG initial conditions changed significantly.
The crucial point to be noted here is that the magnitude 
of the power spectrum with the perturbation is much larger than the spectrum
derived from the MCG initial condition (with $\delta f=0$). For
$\delta f \neq 0$ the  magnitude of power spectrum for even $l$ is   dominant
over its odd counterparts at $T=170$ MeV in contrast to the case with
$\delta f=0$ where the difference in the values of $C_l$ 
between odd and even $l$ is small. 
This is because the perturbation $\delta f$ for $n=2$ has a symmetry under 
the transformation $\theta \leftrightarrow -\theta$ indicating the dominance 
of even $l$ through spherical harmonics. Therefore, dominance
of even $l$ in this particular case indicate the presence of perturbations.
The power spectrum with perturbation at $p_T=1.5$ GeV are displayed in Figs.~\ref{fig10},
~\ref{fig11} and ~\ref{fig12}. The amplitude of the spectrum is larger at all temperatures
compared to the case with $p_T=0.6$ GeV. Qualitatively the spectra is similar to the 
one at lower $p_T= 0.6$ GeV. We find significant enhancement at odd $l$.
\begin{figure}
\centerline{\includegraphics[height=120mm,width=90mm,angle=-90]{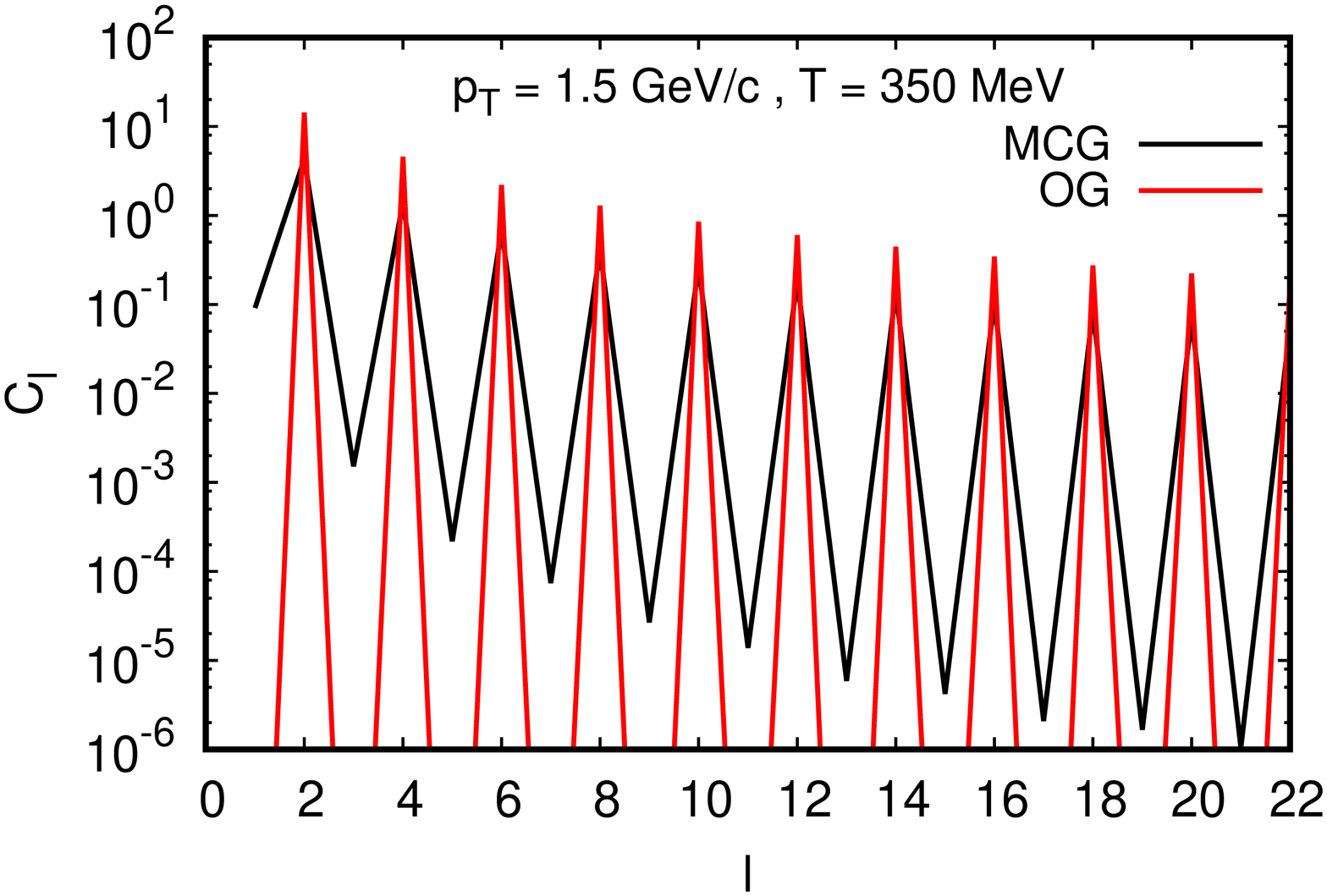}}
\caption{Same as Fig~\ref{fig7} for $p_T = 1.5 \text{ GeV}/c$}
\label{fig10}
\end{figure}
\begin{figure}
\centerline{\includegraphics[height=120mm,width=90mm,angle=-90]{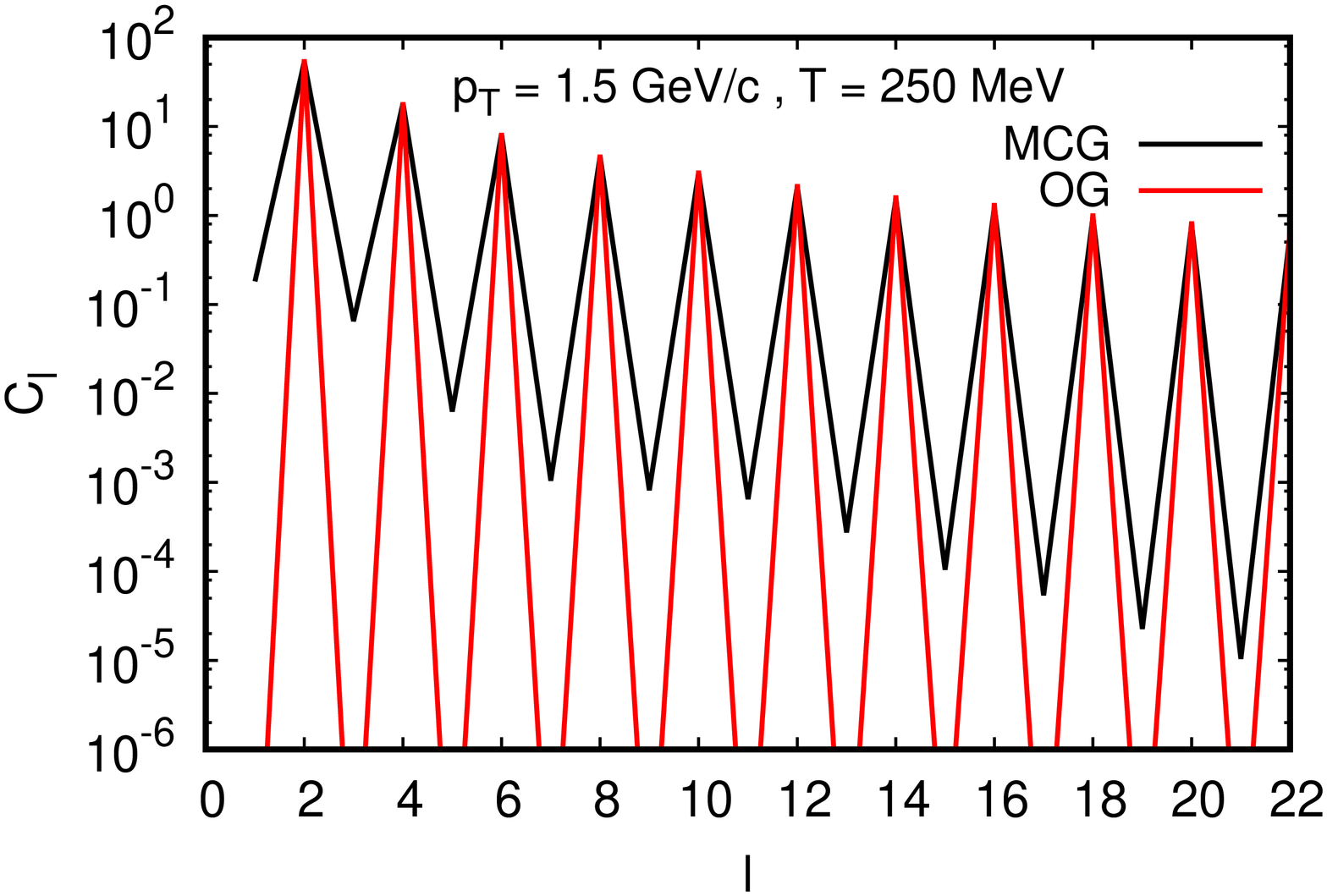}}
\caption{Same as Fig~\ref{fig8} at $T= 170\text{ MeV}$}
\label{fig11}
\end{figure}

The variation of power spectrum with $T$ for different $l$  has been
displayed in Fig.~\ref{fig13}. The $C_l$'s for small $l$ decrease with $T$ monotonically
at low $T$ and reach a plateau at higher $T$. The $C_l$'s for larger $l$ does
not show much variation with $T$. Similar quantities
have been depicted in Fig.~\ref{fig14} for MCG initial condition. 
The  fall is faster in case of MCG initial conditions.  
In Fig.~\ref{fig15} the variation of power spectrum with $T$ for odd $l$
has been depicted. We clearly observe that the $C_l$ falls  faster 
with $T$ as compared to even $l$.
\begin{figure}
\centerline{\includegraphics[height=120mm,width=90mm,angle=-90]{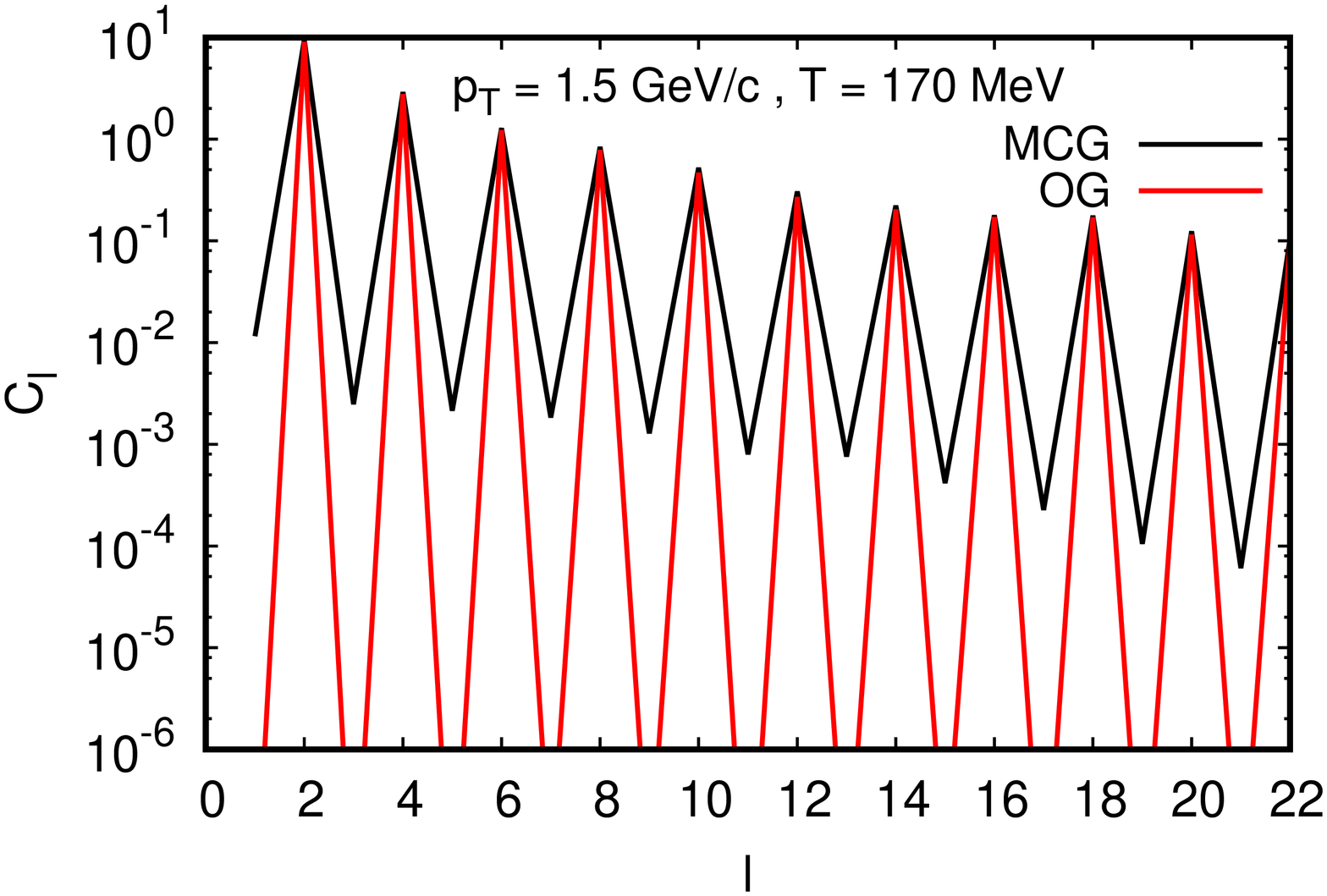}}
\caption{Same as Fig~\ref{fig9} for $p_T = 1.5 \text{ GeV}/c$}
\label{fig12}
\end{figure}
Now we discuss the $T$ variation of power spectrum with inclusion of
perturbation ($\delta f$) for even $l$ (the values with
odd $l$'s are very small). In Fig.~\ref{fig16} (Fig.~\ref{fig17}) the $C_l$ is
plotted as a function of temperature for $p_T=0.6$ GeV for OG (MCG) initial 
condition. We observe the following features: for a given $l$, the value of $C_l$'s are
larger in MCG than OG initial condition, for large $l (\geq 8)$ the $C_l$'s  
do not vary significantly with $T$, most significantly 
the variation with $T$ is non-monotonic. 
Similar studies have been made for higher $p_T=1.5$ GeV.
The results are shown in Figs.~\ref{fig18} and \ref{fig19}. The values of $C_l$'s are much
larger but the qualitative behaviour is same as lower $p_T$ for both the initial conditions.
The important point to be noted here is that the nature of the 
power spectrum on the constant temperature surfaces with perturbation
is distinctly different from the spectrum obtained without perturbation. 
Therefore, observation of the variation of $C_l$ with $T$ will help
in tracing  non-equilibrium features present in the system. 

\begin{figure}
\centerline{\includegraphics[height=120mm,width=90mm,angle=-90]{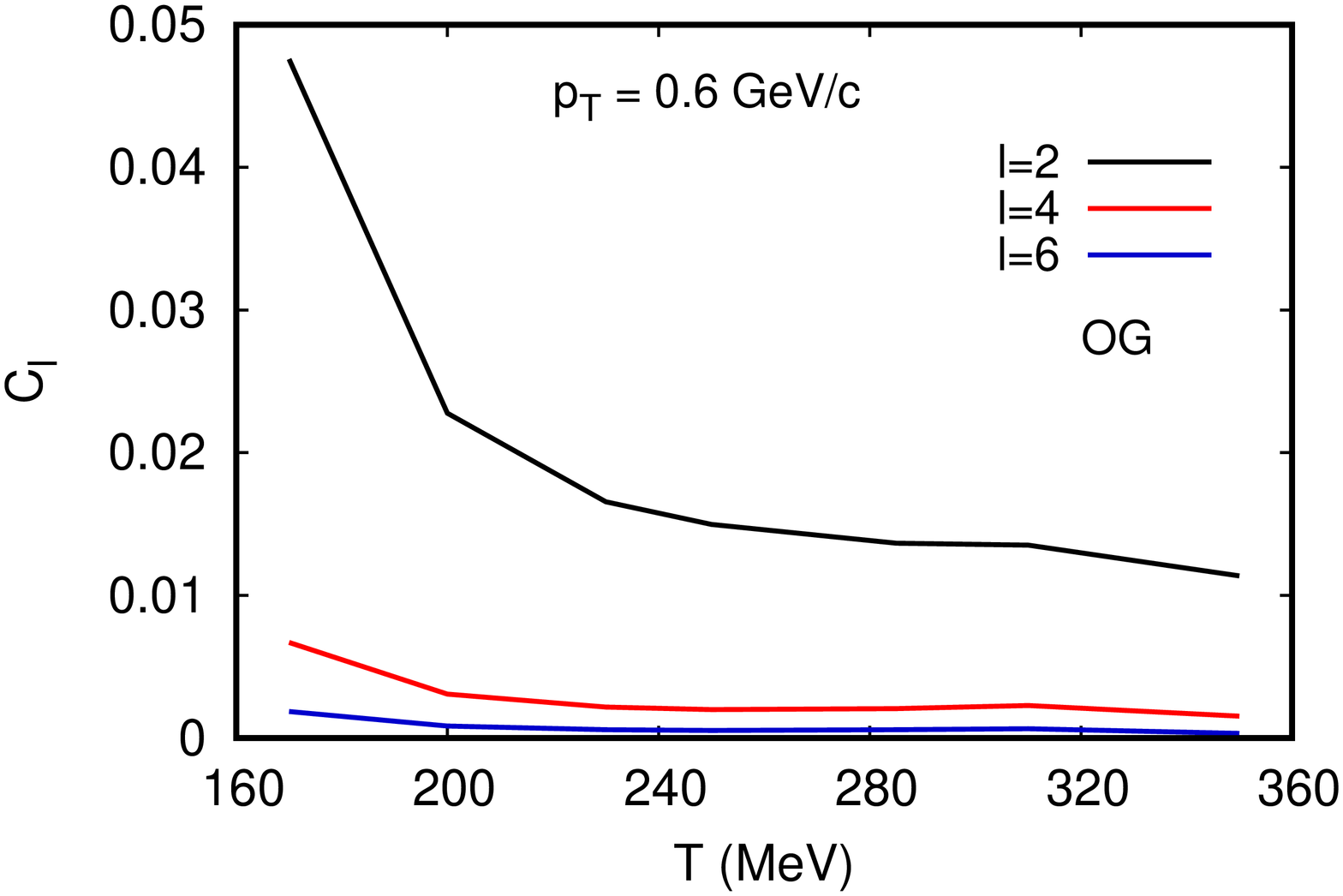}}
\caption{The temperature variation of power spectrum with OG initial 
condition for different $l$ values.}
\label{fig13}
\end{figure}
\subsection{Relation between flow harmonics and power spectrum}
The $\eta$ and $p_T$ dependence of various flow harmonics can be calculated from the power
spectrum using the following relation  (see appendix B for derivation):
$${2\pi}\frac{dN}{p_Tdp_Tdy}v_k=
\sum _{l\ge k}^{\infty}\sqrt{\frac{2l+1}{4\pi}\frac{(l-k)!}{(l+k)!}}
\left[ a_{l,k}+(-1)^ka_{l,-k} \right]P^k_l(\text{cos }\theta)$$
For example, the elliptic flow can be calculated by using values of $P_l^k$ as:
\begin{equation}
v_2=\left(\frac{dN}{2\pi p_Tdp_Tdy}\right)^{-1}\,sech^2\eta\left[a^\prime_{22}+
a^\prime_{32}tanh\eta+a^\prime_{42}(7tanh^2\eta-1)+.....\right]
\end{equation}
where
\begin{equation}
a^\prime_{22}=\frac{1}{3}\sqrt{\frac{5}{12\pi}}a_{22},\,\,
a^\prime_{32}=15\sqrt{\frac{7}{120\pi}}a_{32},\,\,
a^\prime_{42}=\frac{15}{2}\sqrt{\frac{1}{40\pi}}a_{42},
\end{equation}
where $a_{lm}$ is a function of $p_T$ and  given by Eq.~\ref{eqalm}.
The fluctuations in $v_2$ and its dependence on kinematic variables can also be estimated
from the analysis of $Ed\delta N/d^3p$ presented here.
The fluctuations of other harmonics and its dependence 
on $p_T$, $\eta$ can also be calculated using similar procedure.

\begin{figure}
\centerline{\includegraphics[height=120mm,width=90mm,angle=-90]{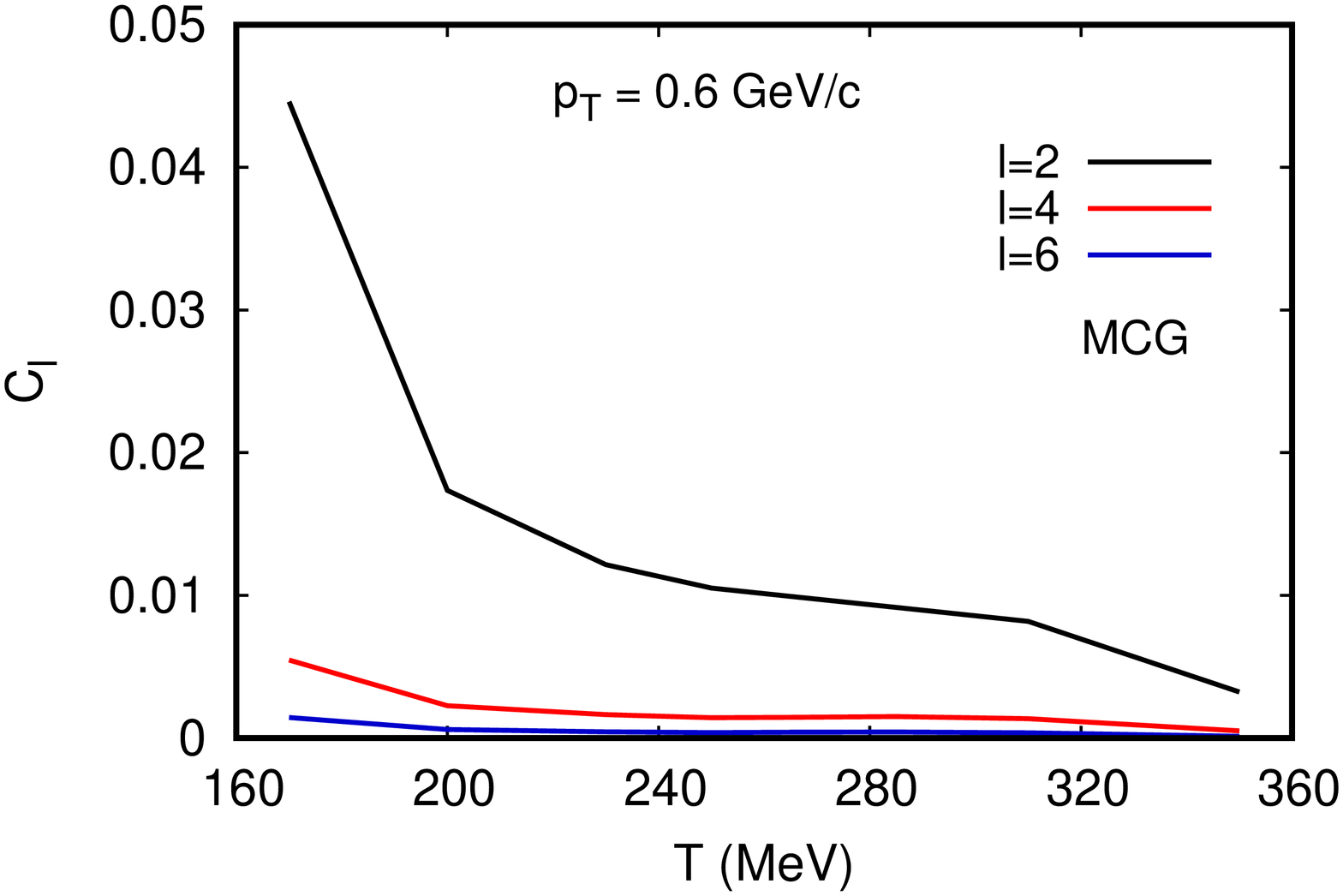}}
\caption{The variation of power spectrum for even $l$ with temperature
for MCG initial condition.
}
\label{fig14}
\end{figure}
\begin{figure}
\centerline{\includegraphics[height=120mm,width=90mm,angle=-90]{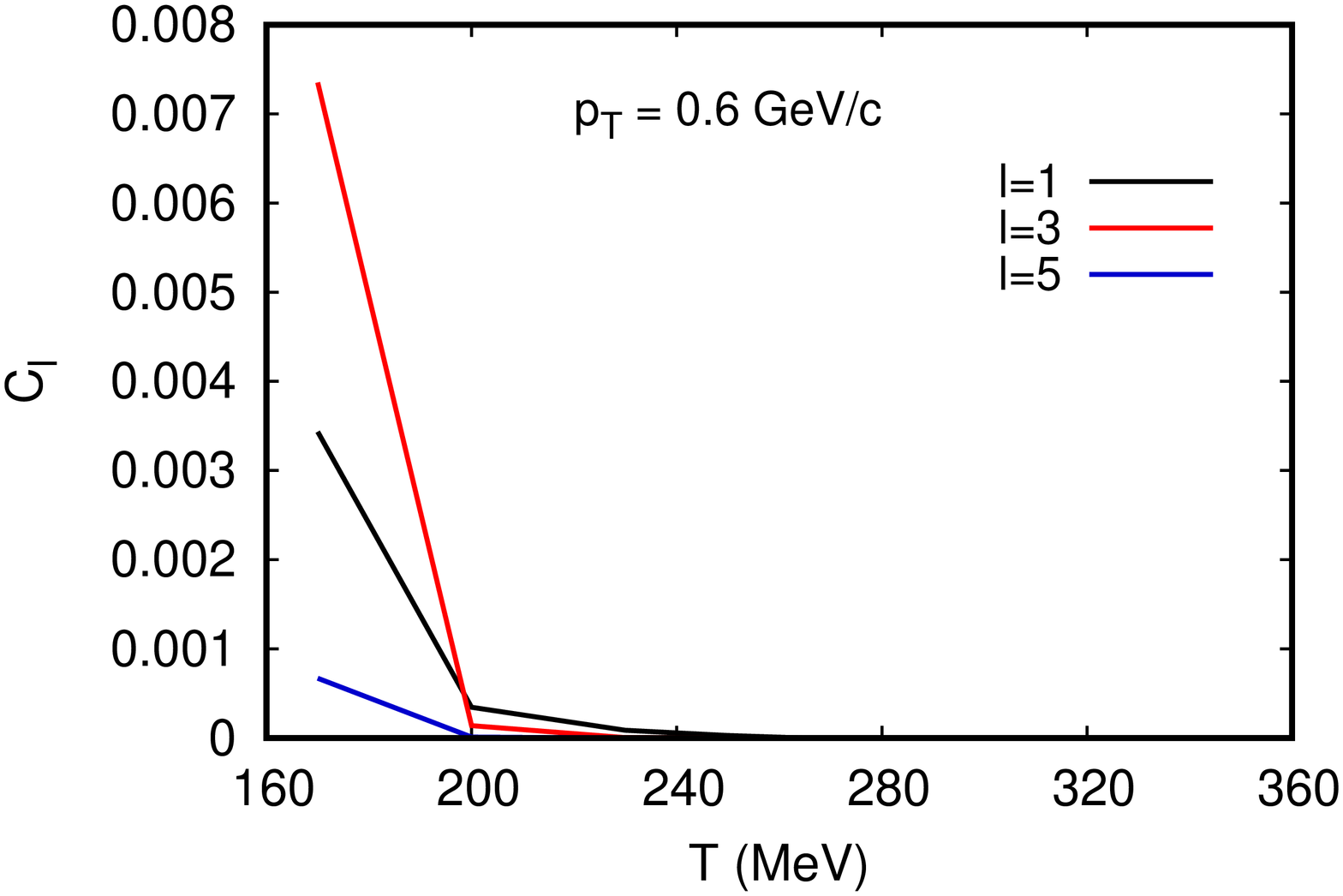}}
\caption{The variation of power spectrum for odd $l$ with temperature
for MCG initial condition.}
\label{fig15}
\end{figure}
\section{Summary and discussions}
The hot and dense system formed in heavy ion collisions at relativistic energies has been evolved
using (3+1) dimensional relativistic hydrodynamics.  The initial energy density profiles 
required to solve the hydrodynamics has been derived from OG and MCG models. 
The power spectrum of momentum distribution of particles  due to fluctuations in 
initial conditions for both OG and MCG models
have been estimated at different surfaces of constant temperatures following the analysis procedure that
is used for CMBR spectrum. This enable us to study the evolution of the power spectrum with 
decrease in temperature and hence effectively with increase in time. 
We observe that the power spectrum with OG initial conditions for central collisions
does not change significantly with the progression of time because the initial system is 
symmetric. However, the nature of the power spectrum 
for MCG initial condition  with negligible values for odd $l$'s changes substantially  
showing similar magnitudes for odd and even $l$'s at later time. 

The power spectrum  for perturbation 
introduced through phase space distribution which derive the system away from equilibrium 
has also been estimated. It has been observed that the temperature variation of
power spectrum with perturbation is distinctly different from the one without
perturbation - clearly indicating the trace of non-equilibration in the system.    
Such studies will help in constraining the initial states~\cite{ERetinsaya,HPetersen}. 
A relation between the power spectrum with the flow harmonics has been derived.
This relation can be used to estimate the pseudo-rapidity and $p_T$  dependence of the
flow harmonics. The power spectrum of phase space perturbation can be used
to estimate the fluctuations in flow harmonics and its dependence on kinematic variables.
The  connection between the experiments and the present type of works  has been
nicely discussed in Ref.~\cite{estrada}. 
\begin{figure}
\centerline{\includegraphics[height=120mm,width=90mm,angle=-90]{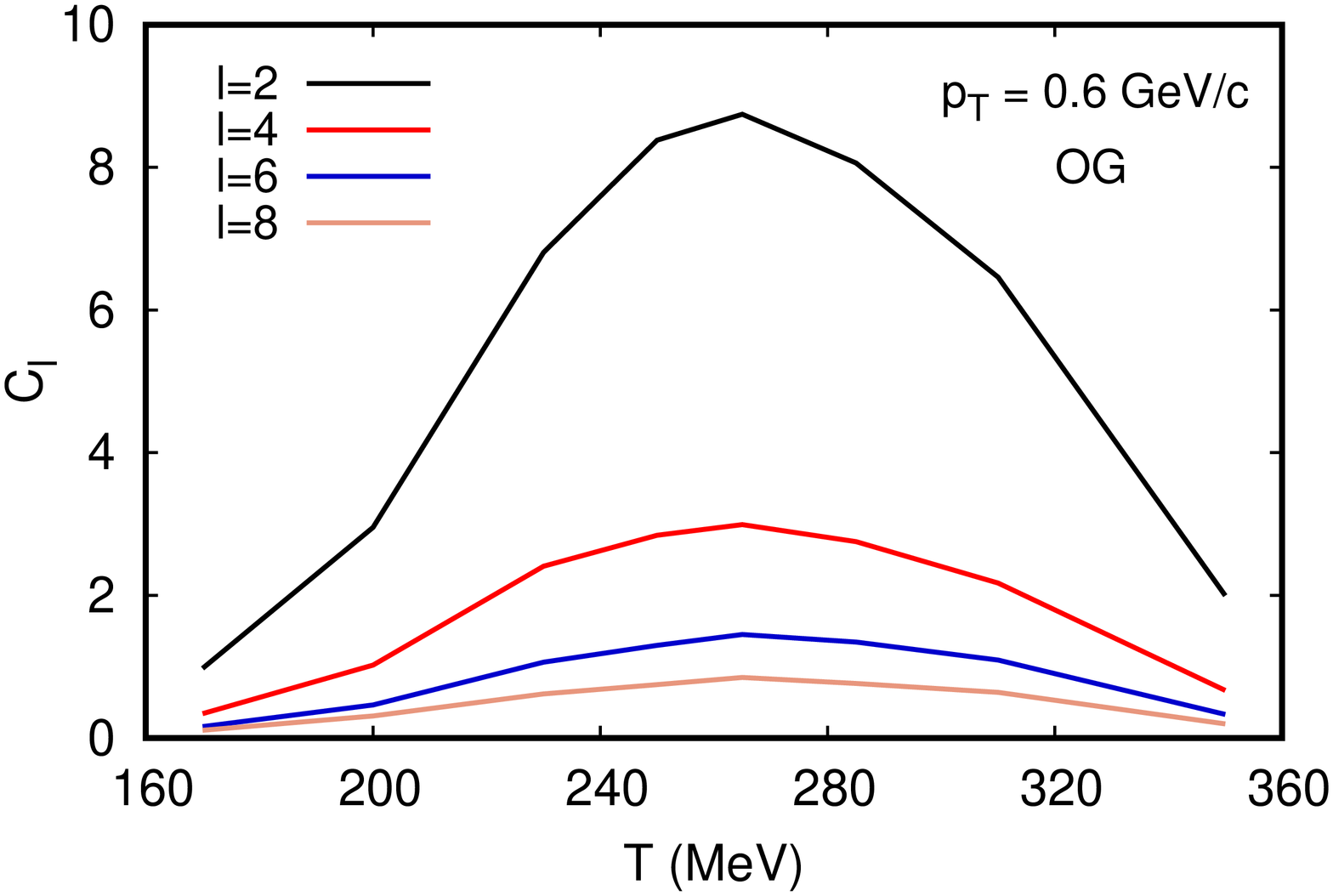}}
\caption{The variation of power spectrum with temperature for OG initial condition 
with perturbation at $p_T = 0.6 \text{ GeV}/c$ (see text for details).}
\label{fig16}
\end{figure}
\begin{figure}
\centerline{\includegraphics[height=120mm,width=90mm,angle=-90]{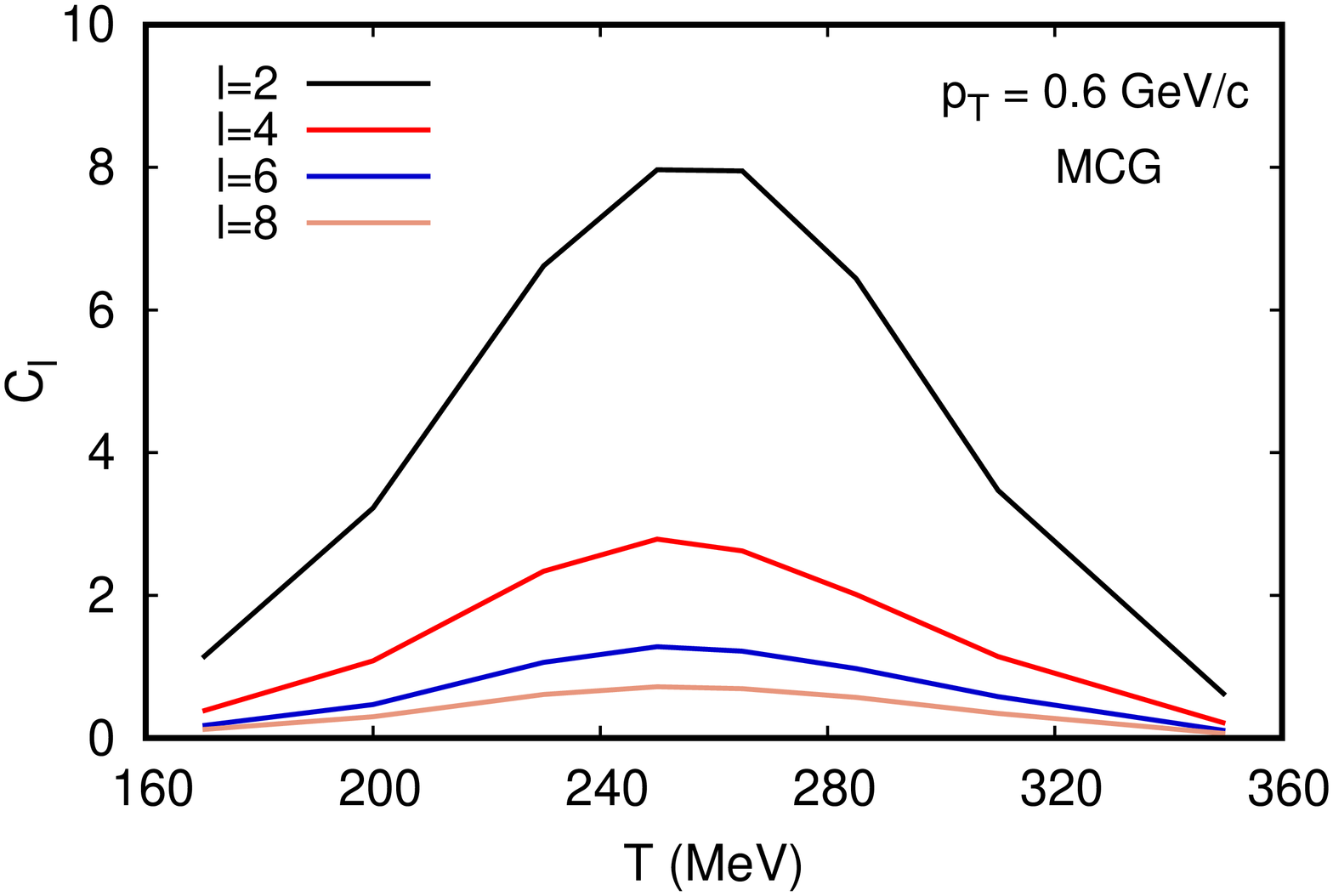}}
\caption{The variation of power spectrum of even $l$ with temperature for MCG 
initial condition with perturbation at $p_T = 0.6 \text{ GeV}/c$ (see text for details).}
\label{fig17}
\end{figure}
\begin{figure}
\centerline{\includegraphics[height=120mm,width=90mm,angle=-90]{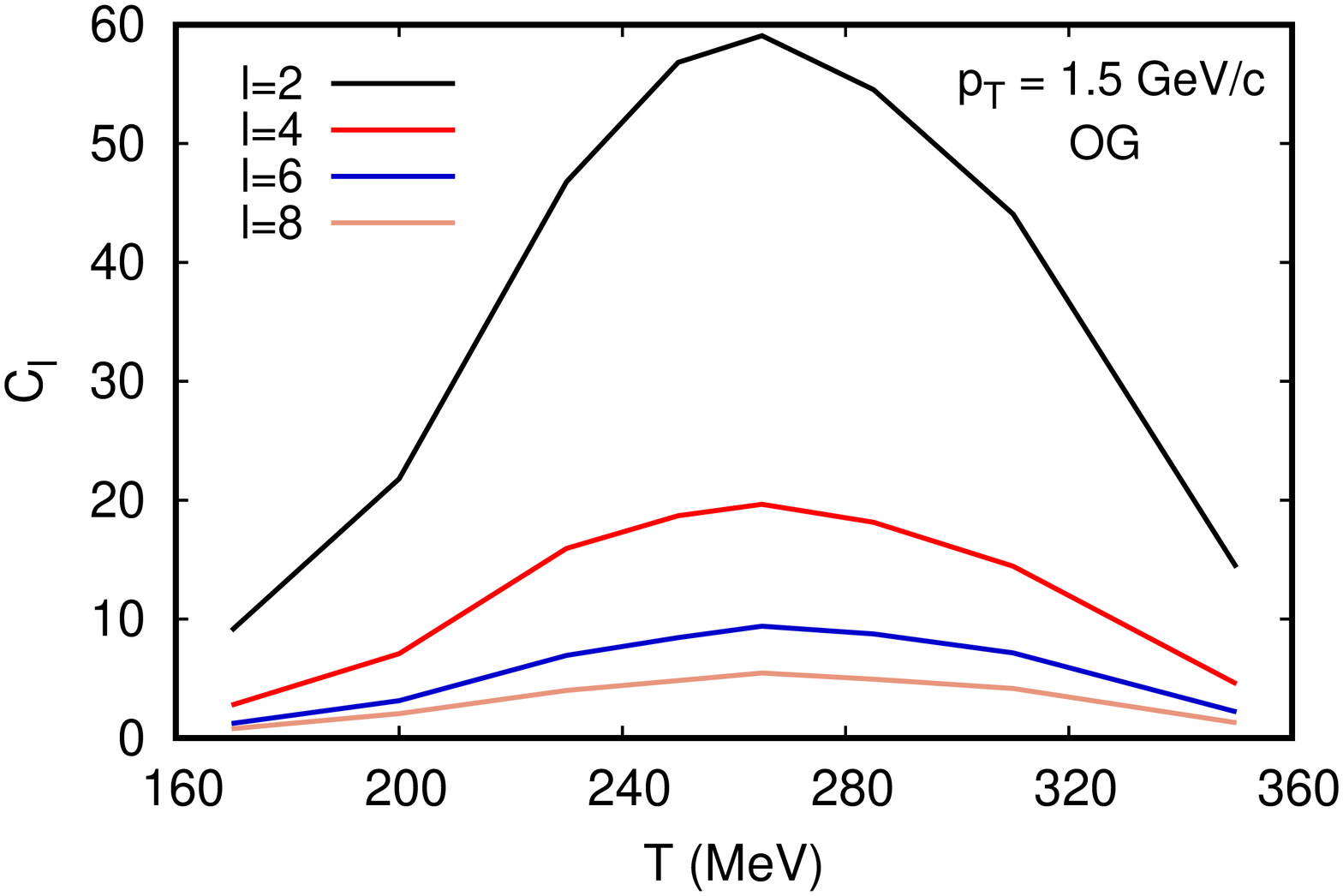}}
\caption{Same as Fig~\ref{fig16} for $p_T = 1.5 \text{ GeV}/c$}
\label{fig18}
\end{figure}
\begin{figure}
\centerline{\includegraphics[height=120mm,width=90mm,angle=-90]{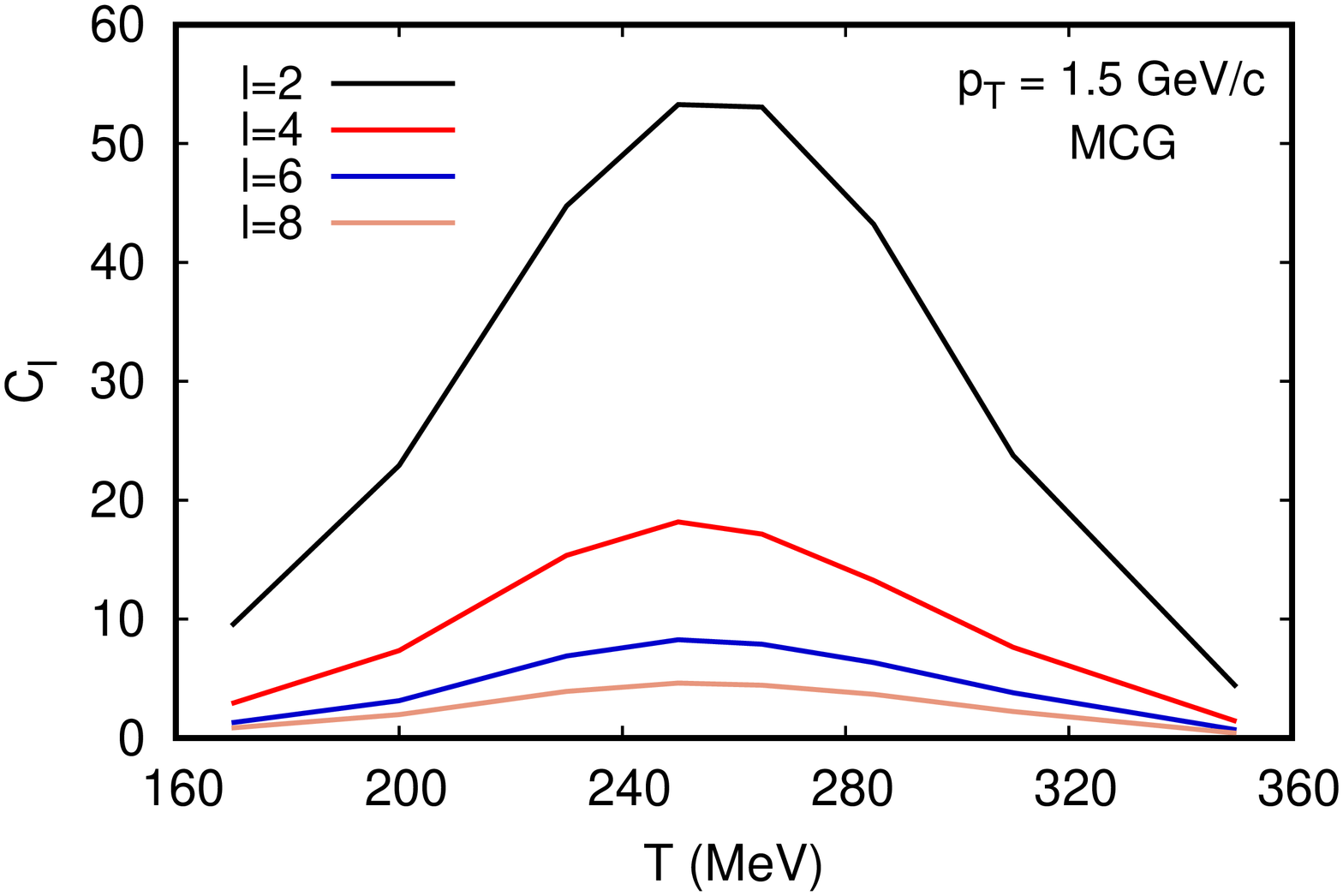}}
\caption{Same as Fig~\ref{fig17} for $p_T = 1.5 \text{ GeV}/c$}
\label{fig19}
\end{figure}

\noindent{\bf Acknowledgement:} G. S. acknowledges the support from Department of Atomic Energy, Govt. of India for
this work.

\section{Appendix A}
In this appendix we provide the expressions for $f_n$'s  appearing in Eq.~\ref{pressure}.
 \begin{equation}
  f_0=1+\frac{3N_f}{32}(7+120 \hat{\mu}_q^2+240 \hat{\mu}_q^4)
 \end{equation}
 \begin{equation}
  f_2=-\frac{15}{4}\left[1+\frac{N_f}{12} (5+72 \hat{\mu}_q^2+144 \hat{\mu}_q^4)\right]
 \end{equation}
 \begin{equation}
  f_3=30\left[1+\frac{N_f}{6} (1+12 \hat{\mu}_q^2)\right]^{3/2}
 \end{equation}
 \begin{equation}
  \begin{split}
 f_4&=237.223+(15.963+124.773 \hat{\mu}_q^2-319.849 \hat{\mu}_q^4)N_f-(0.415+15.926 \hat{\mu}_q^2+106.719 \hat{\mu}_q^4)N_f^2\\
  & + \frac{135}{2}\left[ 1+\frac{N_f}{6} (1+12 \hat{\mu}_q^2)\right]\mbox{ln}
\left[\left(\frac{\alpha_s}{\pi}\right) \left( 1+\frac{N_f}{6} (1+12 \hat{\mu}_q^2)\right)\right]\\
   & -\frac{165}{2}\left[ 1+\frac{N_f}{12} (5+72 \hat{\mu}_q^2+144 \hat{\mu}_q^4)\right]
\left( 1-\frac{2N_f}{33}\right)\mbox{ln }\hat{M}
  \end{split}
 \end{equation}
 \begin{equation}
  \begin{split}
f_5&=-\sqrt{1+\frac{N_f}{6} (1+12 \hat{\mu}_q^2)}[799.149+(21.963-137.33 \hat{\mu}_q^2+482.171 \hat{\mu}_q^4)N_f\\
 & +(1.926+2.0749 \hat{\mu}_q^2-172.07 \hat{\mu}_q^4)N_f^2]+\frac{495}{12}[6+N_f(1
+12 \hat{\mu}_q^2)]\left( 1-\frac{2N_f}{33}\right)\mbox{ln }\hat{M}
  \end{split}
 \end{equation}
 \begin{equation}
  \begin{split}
   f_6&=-\left[ 659.175+(65.888-341.489 \hat{\mu}_q^2+1446.514 \hat{\mu}_q^4)N_f 
 + (7.653+16.225 \hat{\mu}_q^2-516.210 \hat{\mu}_q^4)N_f^2\right. \\
   & \left. -\frac{1485}{2}\left( 1+\frac{N_f}{6} (1+12 \hat{\mu}_q^2)\right)
\left( 1-\frac{2N_f}{33}\right)\mbox{ln }\hat{M} \right]\mbox{ln}\left[ \left(\frac{\alpha_s}{\pi}\right) 
\left( 1+\frac{N_f}{6} (1+12 \hat{\mu}_q^2)\right)4\pi ^2\right]\\
   &-475.587\mbox{ ln}\left[ \left(\frac{\alpha_s}{\pi}\right) 4\pi ^2 C_A\right]
  \end{split}
 \end{equation}

\section{Appendix B}
In this appendix we derive a relation between anisotropic flow coefficients, $v_n$ and the 
coefficients, $a_{lm}$. The $p_T$ distribution can be written as: 
 \begin{equation}
  \frac{dN}{d^2p_Tdy}(p_T,\theta,\phi)=\frac{1}{2\pi}\frac{dN}{p_Tdp_Tdy}\left( 1+\sum _{n=1}^{\infty}v_n\text{cos}(n\phi)\right)
 \end{equation}
which can also be written as:
 \begin{equation}
  \frac{dN}{d^2p_Tdy}(p_T,\theta,\phi)=\bar{N}+\sum _{l=1}^{\infty}\sum_{m=-l}^l a_{lm}Y_{lm}(\theta,\phi)
 \end{equation}
which leads to 
 $$\frac{1}{2\pi}\frac{dN}{p_Tdp_Tdy}\left( 1+\sum _{n=1}^{\infty}v_n\text{cos}(n\phi)\right)=\bar{N}+\sum _{l=1}^{\infty}\sum_{m=-l}^l a_{lm}Y_{lm}(\theta,\phi)$$
 Multiplying both sides of the above Eq. by $\text{cos}(k\phi)$ and integrating  over $\phi$ we get,

 $$\int _0^{2\pi}d\phi \text{ cos}(k\phi)\frac{1}{2\pi}\frac{dN}{p_Tdp_Tdy}\left( 1+\sum _{n=1}^{\infty}v_n\text{cos}(n\phi)\right)=\int _0^{2\pi}d\phi \text{ cos}(k\phi) \left( N_0+\sum _{l=1}^{\infty}\sum_{m=-l}^l a_{lm}Y_{lm}(\theta,\phi)\right)$$ 
 This gives
 $$\frac{1}{2\pi}\frac{dN}{p_Tdp_Tdy}\int _0^{2\pi}d\phi \text{ cos}(k\phi)\sum _{n=1}^{\infty}v_n\text{cos}(n\phi)=\int _0^{2\pi}d\phi \text{ cos}(k\phi) \sum _{l=1}^{\infty}\sum_{m=-l}^l a_{lm}Y_{lm}(\theta,\phi)$$
Now $v_k$ can be expressed as:
 $$\frac{1}{2\pi}\frac{dN}{p_Tdp_Tdy}v_k=\int _0^{2\pi}d\phi \text{ cos}(k\phi) \sum _{l=1}^{\infty}\sum_{m=-l}^l a_{lm}Y_{lm}(\theta,\phi)$$ 
Writing $Y_{lm}$ in terms of associated Legendre polynomials and $e^{im\phi}$ we obtain,
 $$\frac{1}{2\pi}\frac{dN}{p_Tdp_Tdy}v_k=\int _0^{2\pi}d\phi \text{ cos}(k\phi) \sum _{l=1}^{\infty}\sum_{m=-l}^l a_{lm}\sqrt{\frac{2l+1}{4\pi}\frac{(l-m)!}{(l+m)!}}P^m_l(\text{cos }\theta)[\text{cos}(m\phi) +i \text{ sin}(m\phi)]$$  
Performing the $\phi$ integration we get,
 $$\frac{1}{2\pi}\frac{dN}{p_Tdp_Tdy}v_k= \sum _{l\ge k}^{\infty}\sqrt{\frac{2l+1}{4\pi}}\left[ a_{l,k}\sqrt{\frac{(l-k)!}{(l+k)!}}P^k_l(\text{cos }\theta)+a_{l,-k}\sqrt{\frac{(l+k)!}{(l-k)!}}P^{-k}_l(\text{cos }\theta) \right]$$ 
On simplification we obtain the relation between the flow harmonics and $a_{l,k}$ as:
 $$\frac{1}{2\pi}\frac{dN}{p_Tdp_Tdy}v_k= \sum _{l\ge k}^{\infty}\sqrt{\frac{2l+1}{4\pi}\frac{(l-k)!}{(l+k)!}}\left[ a_{l,k}+(-1)^ka_{l,-k} \right]P^k_l(\text{cos }\theta)$$ 



\end{document}